\documentclass[twocolumn]{aastex63}

\usepackage{nicefrac}
\usepackage{savesym}
\savesymbol{tablenum}
\usepackage{siunitx}
\usepackage{amsmath}
\usepackage{mathtools}
\usepackage{booktabs}
\usepackage{multirow}
\usepackage[normalem]{ulem}
\usepackage{natbib}

\restoresymbol{SIX}{tablenum}
\DeclareSIUnit\parsec{pc}
\DeclareSIUnit\years{yr}
\DeclareSIUnit\Msol{M_{\odot}}
\DeclareSIUnit\Lsol{L_{\odot}}
\DeclareSIUnit\AU{au}
\DeclareSIUnit\om{\Omega}
\DeclareSIUnit\orb{T_{\mathrm{orb}}}
\DeclareSIUnit\scaleheight{H}
\definecolor{dodgerblue}{rgb}{0.11764706, 0.56470588, 1.}
\definecolor{seagreen}{rgb}{0.18039216, 0.54509804, 0.34117647}
\definecolor{maroon}{rgb}{0.50196078, 0., 0.}

\shorttitle{VSI with Realistic Thermal Relaxation}
\shortauthors{Pfeil \& Klahr}

\newcommand{\rev}[1]{{{{#1}}}}
\newcommand{\secrev}[1]{{{{#1}}}}

\begin{document}

\title{The Sandwich Mode for Vertical Shear Instability in Protoplanetary Disks}

\author[0000-0002-4171-7302]{Thomas Pfeil}
\affiliation{\centering Max-Planck-Institut f\"ur Astronomie, K\"onigstuhl 17, D-69117 Heidelberg, Germany}
\affiliation{\centering University Observatory, Faculty of Physics, Ludwig-Maximilians-Universität München, Scheinerstr. 1, 81679 Munich, Germany}
\correspondingauthor{Thomas Pfeil}
\email{pfeil@mpia.de}

\author[0000-0002-8227-5467]{Hubert Klahr}
\affiliation{\centering Max-Planck-Institut f\"ur Astronomie, K\"onigstuhl 17, D-69117 Heidelberg, Germany}


\begin{abstract}
Turbulence has a profound impact on the evolution of gas and dust in protoplanetary disks
(PPDs), from driving the collisions and the diffusion of dust grains, to the concentration of
pebbles in giant vortices, thus, facilitating planetesimal formation. The Vertical Shear Instability (VSI) is a hydrodynamic mechanism, operating
in PPDs if the local rate of thermal relaxation is high enough. Previous studies of the VSI
have, however, relied on the assumption of constant cooling rates, or neglected the finite coupling time between the gas
particles and the dust grains. Here, we present the results of hydrodynamic
simulations of PPDs with the PLUTO code that include a more realistic 
thermal relaxation prescription, which enables us to study the VSI in the optically thick
and optically thin parts of the disk under consideration of the thermal dust-gas coupling. We show the
VSI to cause turbulence even in the optically thick inner regions of PPDs in our two- and
three-dimensional simulations. The collisional decoupling of dust and gas particles
in the upper atmosphere and the correspondingly inefficient thermal relaxation rates lead to the
damping of the VSI turbulence. Long-lived
anticyclonic vortices form in our three-dimensional simulation. These structures emerge from the
turbulence in the VSI-active layer, persist over hundreds of orbits and extend vertically over
\rev{the whole extent of the turbulent region}. We conclude that the VSI leads to turbulence and the formation of
long-lived dust traps \rev{within $\pm 3$ pressure scale heights distance from the disk midplane}.
\end{abstract}

\keywords{protoplanetary disks --- accretion, accretion disks --- hydrodynamics --- instabilities  --- methods: numerical}

\section{Introduction} \label{sec:intro}
Turbulence plays an important role in the evolution of protoplanetary disks (PPDs) \citep{Weizsaecker1943, Shakura1973, LyndenBell1974, Pringle1981} and in the coagulation and diffusion of dust, and the formation of planetesimals within them \citep{Voelk1980, Ormel2007, Johansen2014, Ishihara2018, Gerbig2020, Klahr2020}. On small scales, it is responsible for the turbulent diffusion of solids, and therefore counteracts the formation of dense clumps \citep{YoudinLithwick2007}. On larger scales, it can trigger the formation of flow structures like zonal flows and anticyclonic vortices \citep{Klahr2003, Lyra2014, Manger2018, Manger2020} that can accumulate dust and possibly seed streaming instability and facilitate planetesimal formation \citep{Johansen2007, Gerbig2020}. 
In combination with magnetic disk winds \citep{Koenigl1993, Bai2013, Rodenkirch2020}, it is believed to regulate the disk's accretion rate \citep{Shakura1973, LyndenBell1974, Pringle1981} and it is one of the controlling parameters for the large scale distribution of dust \citep{Weidenschilling1977, Birnstiel2009, Flock2017,Lin2019, Flock2020}. But despite an increasing amount of research conducted to find the origin of accretion disk turbulence, it is still unclear what turbulence creating mechanism prevails in which parts of PPDs. 

In  contrast to accretion disks around massive compact objects, circumstellar disks are cold and poorly ionized \citep{Dzyurkevich2013}, \rev{which allows the magnetic fields to diffuse with respect to the gas, i.e.\ one has to consider the equations of non-ideal
magnetohydrodynamics. Magnetorotational Instability \citep[MRI, ][]{Balbus1991} can still operate as long as the diffusion time for magnetic fields is longer than the growth time of the MRI, which is typically on the order of the dynamical time scale of the disk. As the diffusion time depends on the considered length scales, it is possible that certain large scales show magnetohydrodynamic effects and even the growth of MRI modes, yet small scales can perfectly decouple from the magnetic fields and be described by the equations of hydrodynamics. In other terms, the magnetic Reynolds number is by orders of magnitude smaller than the hydrodynamic Reynolds number \citep{Fromang2007a, Fromang2007b, Lyra2011}.}

Purely hydrodynamic sources of turbulence \rev{can either be important if the disk is completely decoupled from the magnetic fields (dead zone), or what is equally interesting, if the scales on which the hydrodynamic instabilities do operate are decoupled from the magnetic field.
Mechanisms like the Vertical Shear Instability \citep{Urpin1998, Nelson2013, Stoll2014, Flock2017, Richard2016, Manger2018, Flock2020, Manger2020} have thus come into the focus of research in the past years \citep{Klahr2018, Lyra2018}.} 

PPDs with a radial gradient in temperature typically have a vertical shear in their azimuthal velocity, as can be deduced from radial hydrostatic balance, i.e.
\begin{equation}
\label{eq:RHE}
\Omega^2 R =\frac{1}{\rho}\frac{\partial P}{\partial R} +\frac{GM_*R}{(R^2+z^2)^{3/2}},
\end{equation} 
where $\Omega$ is the gas' angular frequency, $\rho$ and $P$ are the gas density and pressure respectively, $G$ denotes the gravitational constant, $M_*$ is the central star's mass, and $R$ and $z$ are the radial and vertical coordinate. This circumstance allows gas parcels to conserve their angular momentum while moving vertically and radially in the disk, thus, leading to the violation of Rayleigh's stability criterion for circular shear flows \citep{Drazin2004}. 
The resulting instability is the VSI, which drives turbulence with a strength of $\alpha\sim 10^{-6} - 10^{-3}$. 
In recent years, numerical studies of isothermal disks by \citet{Richard2016} and \citet{Manger2018} have shown the VSI's potential to trigger the formation of giant long-lived anticyclonic vortices. In rotating fluids like PPDs, these structures are common features which have been studies extensively in the past \citep{Goodman1987, Adams1995, Godon1999, Klahr2003, Barranco2005, Meheut2010, Raettig2012, Surville2015}. In their cores, they \rev{produce} a high-pressure region in which inward drifting dust particles can accumulate \secrev{\citep{Whipple1972, Barge1995, Adams1995, Tanga1996, Lyra2013}}. Thus, by creating local dust over-densities, vortices are regions that are particularly suitable to the formation of planetesimals -- the building blocks of planets \citep{Barge1995}. The VSI's ability to form these structures under ideal \rev{(i.e. isothermal, and thus buoyancy-free)} conditions makes it a very interesting mechanism for studies in further refined numerical simulations.

\rev{Fast thermal relaxation is required for VSI to overcome buoyancy forces. Thermal relaxation in stably stratified disks leads to the damping of the internal gravity waves. The maximal damping occurs, when the oscillation period equals the relaxation time. For very long cooling times, there is effectively no damping (adiabatic case), but for the instantaneous cooling the effective buoyancy frequency is zero, as in the isothermal case. In this case there is no restoring force to drive the oscillation. This means that with decreasing cooling times, the stabilizing effect of buoyancy decreases to a level at which the VSI unstable modes can grow, leading to turbulence \citep{Lin2015}.} 

PPDs \rev{have} a broad variety of thermal relaxation regimes, which means there exist regions that are optically thin or optically thick, as well as transition regions of opacity, e.g. at the water ice line. 
Additionally, thermal coupling between the dust and the gas plays an important role because the molecular hydrogen, of which the disk is mostly composed of, can only cool efficiently if its thermal energy is transferred to an emitting species. 
\rev{Note, that optically thick refers to the integral over $\kappa\, \rho\, \mathrm{d}z$, executed over the disk's complete thickness. This does not automatically imply thermal relaxation times longer than a small fraction of an orbital period, neither does optically thin imply cooling times much shorter than $\Omega^{-1}$. In the first case the cooling over an unstable wave length can still be very short, even if it is embedded in an optically thick region of the disk. On the other hand, even in optically thin regions, cooling will be limited by the opacity of the dust and also by the coupling of dust and gas via collisions \citep{Pfeil2019}.}
In the cold regions of the PPD midplane beyond the water ice line, the gas is mostly cooled via the dust grains \citep{Malygin2017}\rev{, while above the midplane and close to the star also gas opacities, especially of evaporated water can become important in some cases \citep{Freedman2008}}.

In regions of slow thermal relaxation, VSI turbulence might be weak or completely suppressed. Other instability mechanisms, operational with lower rates of cooling, could potentially cause turbulence in these zones. \citet{Marcus2016} discussed the Zombie Vortex Instability \rev{\citep[ZVI,][]{Marcus2015}}, as a possible source of turbulence in stably stratified regions of adiabatic gas. Optically thin cooling or radiative diffusion inhibit the ZVI, making it operable in regions where the VSI can not create turbulence.

Another alternative source of turbulence in PPDs with slower cooling and a negative radial entropy gradient is the Convective Overstability \rev{\citep[COS, ][]{Klahr2014}}. This mechanism creates weak turbulence in regions where the thermal relaxation time is $\tau \sim 1/ \gamma \Omega$. COS might thus be active in parts of PPDs that are also susceptible to VSI or have too low cooling rates for the VSI, depending on the local stratification \citep{Pfeil2019}. 

So far, numerical studies of the VSI have relied on isothermal or isentropic equations of state \citep[e.g.][]{Nelson2013}, on a simplified treatment \rev{of radiation hydrodynamics via flux-limited diffusion \citep{Stoll2014,Flock2017} or a spatially and temporally fixed prescribed cooling time \citep{Manger2018}}. It is, thus, not clear to date, how VSI turbulence evolves in PPDs with a complex density and temperature structure, where the local rate of thermal relaxation has a complex spatial distribution dependent on the disk's local stratification.

For this reason, we conduct two- and three-dimensional simulations of stratified circumstellar disks, including a prescription of thermal relaxation that is deduced from the local disk structure and takes processes like collisional dust-to-gas coupling, optically thin and optically thick radiative cooling into account, \rev{which allows for a more realistic study of the VSI in the upper disk atmosphere}. We first introduce our slightly simplified version of the thermal relaxation model by \citet{Malygin2017} in \autoref{sec:Malygin}.

We then deduce the structure of a PPD in centrifugal and thermal equilibrium from the steady state accretion disk model we already used in \citet{Pfeil2019}, to get a set of initial conditions for our simulations (\autoref{sec:Structure}).

The PLUTO code \citep{Mignone2007} and our modifications are introduced in \autoref{sec:Methods}.

To investigate how the VSI turbulence depends on disk stratification and the new cooling times, we conduct a series of simulations for a set of parameters like the radiative diffusion length scale and the global temperature stratification in a two-dimensional setup in \autoref{sec:WavenumberStudy} and \autoref{sec:StratificationStudy}. 

Furthermore, we are interested in how the VSI can form flow structures in a three-dimensional simulation with realistic thermal relaxation. Our results on vortex formation, structure, and evolution are presented in \autoref{sec:3D}.

Finally, we discuss our results in \autoref{sec:Discussion} and give a conclusion and outlook in \autoref{sec:Conclusion}.

\section{Theoretical Background}
\subsection{Thermal Relaxation Model by Malygin et al.}
\label{sec:Malygin}
Following \citet{Malygin2017}, three ways of energy transfer in PPDs are important for the thermal relaxation of linear temperature perturbations. In general, thermal relaxation can only happen via the emission of radiation because thermal conduction is a negligible effect in the dilute gas of PPDs. Thus, to equilibrate an excess/lack of thermal energy with its surrounding, a gas parcel has to emit/absorb radiation via the emitting/absorbing components of the dust and gas mixture. In PPDs, both the dust grains and some emitting gas species contribute in this process. We ignore the gas as a coolant in this study, as discussed later, and only consider the optically thin emission timescale of the dust 
\begin{equation}
    \tau_{\text{emit}}=\frac{C_V}{16 \kappa \sigma_{\text{SB}} T^3},
\end{equation}
where $C_V$ is its specific heat capacity at constant volume, $\kappa$ is its opacity (calculated following \citet{Bell1994}), $\sigma_{\text{SB}}$ is the Stefan Boltzmann constant, and $T$ is the dust's temperature.
This process can only be efficient, if the grains receive the gas' thermal energy via collisions. If these collisions are scarce, the dominating timescale is set by the collision timescale between dust and gas particles
\begin{equation}
    \tau_{\text{coll}}=\frac{1}{n \sigma_{\text{coll}} v_{\text{coll}}},
\end{equation}
where $n$ is the number density of dust grains, $\sigma_{\text{coll}}$ is the collisional cross section of dust and gas particles ($\approx \SI{1.5e-9}{\square \centi \meter}$), and $v_{\text{coll}}$ is the typical collision velocity.
As long as the disk is optically thin and cool, thermal relaxation happens on the timescale
\begin{equation}
    \tau_{\text{thin}} = \max (\tau_{\text{emit}}, \tau_{\text{coll}}).
\end{equation}
However, in the disk's deep interior, close to the midplane, the optical depth can be very high. Thus, radiative diffusion is the dominant transfer process of thermal energy. The corresponding timescale of radiative diffusion is dependent on the physical size of a temperature perturbation $\lambda$, represented by the perturbation wavenumber $k=2\pi/\lambda$. Analysis of the perturbed energy equations by \citet{Malygin2017} results in
\begin{equation}
\label{eq:thick}
	\tau_{\text{diff}}=\frac{1}{f D_E k^2}=\frac{1}{\tilde{D}_E k^2},
\end{equation} 
with the \rev{diffusion coefficient $D_E=\xi c/\kappa \rho$, the flux limiter $\xi$ \citep{Levermore1981}, gas density $\rho$, speed of light $c$,} and the factor $f$, given by
\begin{equation*}
f=\frac{4\eta}{1+3\eta} \qquad \eta=\frac{E_{\text{rad}}}{E_{\text{rad}}+E_{\text{int}}},
\end{equation*}
where $E_{\text{rad}}$ is the radiation energy density, and $E_{\text{int}}$ is the internal energy density of the gas.

The slowest channel of energy transfer limits the total relaxation time to the value
\begin{equation}\label{eq:trelax}
	\tau_{\text{relax}}=\max(\tau_{\text{coll}},\tau_{\text{diff}},\tau_{\text{emit}}).
\end{equation}
The deep interior of the disk is thus dominated by the diffusion timescale, while the upper layers are dominated by the optically thin relaxation timescale. 
\citet{Malygin2017} point out that in the upper layers, photoelectric heating due to stellar irradiation, photochemistry, and the mutual irradiation of the dust and gas particles must be accounted for. As for their approach, we ignore these physical phenomena for the moment and only consider radiative diffusion, the optically thin dust emission, and the collisional coupling of dust and gas for our cooling time prescription that follows \autoref{eq:trelax}. We also assume that in the investigated region, the gas opacity is generally negligible in the upper atmosphere, due to the vertically decreasing temperature. As can be seen in \autoref{fig:Map}, the disk is warmer in the midplane due to viscous heating. In the upper layers, one finds lower temperatures, well below the evaporation temperature of the water ice. Thus, water can assumed to be in solid form in the upper parts of our simulation domain, meaning the respective opacity is low \citep{Freedman2008} and thermal relaxation in the atmosphere is dominated by the dust grains' emission \citep{Malygin2017}.

\subsection{Structure and Stability of Protoplanetary Disks}
\label{sec:Structure}
In \citet{Pfeil2019}, we investigated the structure of PPDs under consideration of viscous heating, stellar irradiation, and hydrostatic and thermal balance. From these studies, we were able to map where in PPDs certain instability mechanisms, like the VSI, could potentially operate. 
Here, we use the same methods to model the disk structure and stability of the interior parts of a PPD with realistic radial and vertical stratification. The resulting midplane density and temperature structure is used as the initial condition for our hydrodynamic simulations, presented in the next sections. 

\begin{figure}[ht]
    \centering
    \includegraphics[width=0.45\textwidth]{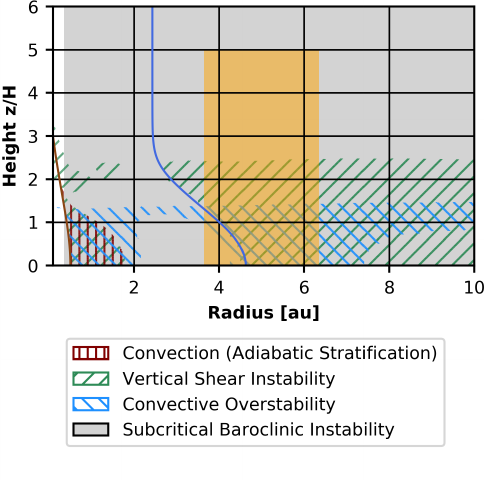}
    \caption{Stability map of a PPD around a T-Tauri star, showing where the disk is susceptible to the different hydrodynamic instabilities \citep[see][]{Pfeil2019}. \rev{The orange} region is the part of the disk around the water ice line that we \rev{intend to study with our 2D and 3D simulations. In our simulations with thermal relaxation instead of full radiation transport, we simplify the density and temperature structure by radial power laws and a vertically isothermal structure, i.e.\ $\beta_T=-1$, $\beta_\rho=-1.5$} (our domain spans $\pm 5$ pressure scale heights vertically).}
    \label{fig:Map}
\end{figure}

\begin{figure*}[ht]
    \centering    \includegraphics[width=\textwidth]{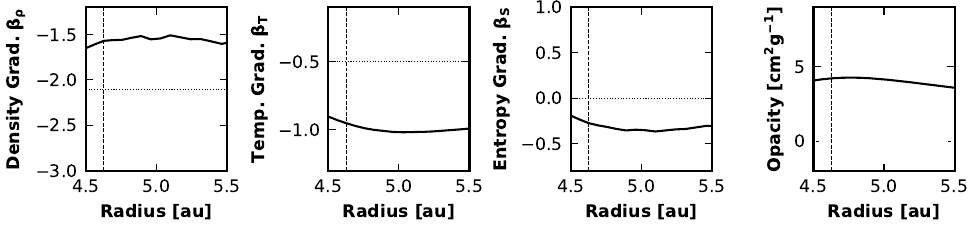}
    \caption{Local radial midplane stratification of the PPD that we simulate. We approximate $\beta_T=-1$ and $\beta_\rho=-1.5$ for the whole simulation domain.}
    \label{fig:Gradients}
\end{figure*}

We model the structure of a disk orbiting a solar-mass T-Tauri star \rev{\citep[stellar parameters as obtained by][]{Baraffe2015}}, with viscosity parameter $\alpha=5\times 10^{-3}$, a mass accretion rate of \SI{5e-8}{\Msol \per \years}, and a disk mass of $M_{\text{disk}}= \SI{0.0823}{\Msol}$. From the obtained disk structure, we created a stability map, similar to those presented in \citet{Pfeil2019}, which is shown in \autoref{fig:Map}. For our studies of the VSI we want to capture different regimes of thermal relaxation of the disk and chose a region around the midplane water ice line. \autoref{fig:Gradients} depicts the respective local structure. At this location, the radial stratification in temperature is comparably steep ($\beta_T=-1$), the opacities are high, and the considered atmospheric layers are cool. This allows us to study the instability under non-ideal conditions \rev{(i.e. a non-isothermal gas with finite cooling time)} close to the midplane, where radiative diffusion is the dominant cooling process, and in the dilute atmospheric layer, where in our case the collisional timescale determines the thermal relaxation rate.

In \autoref{fig:Map}, it can be seen that the conditions at this location, marked by the orange rectangle, are sufficient for the VSI for a height of up to $\sim 2-3$ pressure scale heights. Higher up in the atmosphere, collisional decoupling of the dust and gas components makes cooling very inefficient, which is prohibitive for the VSI.
At closer distance to the central star, the density and, thus, the optical depth become too large for the VSI and the instability is quenched in most parts of the very inner disk. Inside of $\approx \SI{1}{\AU}$, viscous heating and high optical depth lead to a vertically adiabatic structure, i.e. convection can occur. Under such conditions, the VSI is again able to operate, even for very weak cooling \rev{\citep{Nelson2013, Lin2015, Pfeil2019}}. Simulations with a vertically adiabatic stratification are appropriate for the investigation of the VSI in such regions. 
Note, that the disk's vertical stratification in the region we are interested in, is also far from vertically isothermal close to the midplane, as can be seen at the bend water ice-line in \autoref{fig:Map}. The reason for this is the viscous heating, that is assumed to operate in our disk structure model.
The simulations of the marked region, presented here, are, however, vertically isothermal. A vertically isothermal disk is in fact less susceptible to the VSI than a disk with a negative vertical temperature gradient, as obtained from our disk structure model. The reason for this is that a vertically perturbed gas parcel is subject to buoyancy forces, which are weaker if the gas' temperature is decreasing with height, such that the rising gas bubble can equilibrate faster than in an isothermal disk.

From \autoref{fig:Map}, it becomes evident that the simulated region is also susceptible to the COS, and the Subcritical Baroclinic Instability \citep[SBI,][]{Klahr2003, Petersen2007a, Petersen2007b}. However, since the VSI's growth rate is usually much higher under the conditions we investigate, we expect the VSI to be the dominant mechanism in our simulations. The SBI should be considered to operate as an additional mechanism for the enhancement and stabilization of large-scale anticyclonic vortices in three-dimensional setups.

\section{Method}
\label{sec:Methods}
We use the PLUTO\footnote{\url{http://plutocode.ph.unito.it/}} code to solve the equations of inviscid hydrodynamics in our simulations.
The Euler equations, solved by PLUTO read
\begin{align}
\frac{\partial \rho}{\partial t}+\vec{\nabla}\cdot (\rho\vec{v})&=0 \\
\frac{\partial \rho \vec{v}}{\partial t}+\vec{\nabla}\cdot (\rho\, \vec{v}\, \vec{v}^\text{T})&=-\vec{\nabla}P - \rho \vec{\nabla}\Phi, 
\end{align}
where $\rho$ is the gas density, $\vec{v}$ is the gas velocity vector, $P$ denoted the pressure, and $\Phi$ is the gravitational potential.
The ideal equation of state is used as a closure relation, i.e.
\begin{equation}
   P=\frac{k_\text{B} T}{\mu m_{\text{p}}}\rho, 
\end{equation}
with $k_\text{B}$ being Boltzmann's constant, $\mu m_{\text{P}}=2.33\times \SI{1.67e-24}{\gram}$, being the mean molecular mass of the gas, and the gas temperature $T$.
PLUTO provides several methods for solving this set of equations. For our purpose, we chose the combination of the HLLC Riemann solver \citep{Toro1994}, the WENO3 reconstruction scheme \citep{Yamaleev2009} and the third-order un-split Runge-Kutta time integrator. In our two-dimensional simulations, we set the CFL number to 0.4, while in three dimensions, 0.3 is chosen \citep[see][]{Beckers1992}. This combination has very little numerical diffusion and runs reliably stable over the desired simulated timescale.

To realize cooling on the timescale described in the previous section, we employ simple Newtonian cooling, similar to the methods used by \citet{Nelson2013} or \citet{Manger2018}
\begin{align}
    \frac{\text{d} P}{\text{d}t} &= - \frac{P-P_{\text{a}}}{\tau_{\text{relax}}} \\
    \Rightarrow P &= P_a + (P_0-P_a)\exp\left(\frac{-\Delta t}{\tau_{\text{relax}}}\right),
\end{align}
where $P_0$ is the pressure within the grid cell before the cooling, $P_a$ is the desired new pressure i.e. the local pressure of the initial condition, and $\tau_{\text{relax}}$ is the relaxation time, calculated following the previous section. $\tau_{\text{relax}}$ is usually expressed via the parameter $\beta$ in units of the local orbital timescale $\beta\coloneqq \tau_{\text{relax}}\Omega$. Because of this, Newtonian cooling is sometimes referred to as $\beta$-cooling.
We chose the opacity model by \citet{Bell1994} to calculate the relaxation times, following \autoref{eq:trelax}. 

A steep radial stratification in temperature, as employed in our simulations, typically occurs where high opacities lead to a heat build-up caused by viscous heating in the midplane. Choosing a steep stratification, thus comes of the cost of very fast cooling if one wants to include realistic thermal relaxation.
Due to the fact that the calculation of the radiative diffusion timescale requires the assumption of a diffusion length scale (see \autoref{eq:thick}), we introduce the wavenumber $kH$ as a new parameter, where $H$ is the disk's pressure scale height. In our simulations, $kH$ is held at a constant pre-defined value throughout the entire duration of a run. It is therefore not a representation of the real size of any temperature perturbation in our simulations, but just a parameter that sets the efficiency of thermal relaxation within the optically thick parts of the disk. We thus study a broad range of $kH$ values in the later sections, to assess the sensitivity of our simulation results to this parameter.

\subsection{Simulation Setup}

For the simulations of the local patch of a PPD, marked by the orange box in \autoref{fig:Map}, we set up an equilibrium density and temperature structure in the radial-polar plane (spherical coordinates) in code units (cu), following
\begin{align}
\frac{c_s}{v_0}&=h_0 R^{\beta_T/2} \\
\frac{H}{L_0}&= \frac{c_s}{\Omega} = c_s R^{(3+\beta_T)/2} \\
\frac{\rho}{\rho_0}&= R^{\beta_{\rho}} \exp\left(\frac{R^2}{H^2} \left(\frac{R}{\sqrt{R^2+z^2)}} - 1\right)\right) \\
\frac{P}{P_0}&=\rho c_s^2,
\end{align}
with $c_s$ being the speed of sound, $h_0=H_0/L_0$ being the disk's aspect ratio at the reference distance $L_0$, $H=c_s/\Omega$ being the pressure scale height, and the cylindrical radial and vertical coordinates $R,z$. The steepness of the radial power laws in density and temperature in the disk midplane is given by the exponents $\beta_{\rho} = \mathrm{d} \ln(\rho) / \mathrm{d}\ln(R)$ and $\beta_{T} = \mathrm{d} \ln(T) / \mathrm{d}\ln(R)$.
The initial velocities are set to $v_r=v_{\theta}=0$ and
\begin{align}
\frac{v_{\phi}}{v_0} &= \sqrt{\frac{1}{R}}\sqrt{1+\beta_{T} - \frac{\beta_{T}R}{\sqrt{R^2+z^2}} + \rev{(\beta_T+\beta_{\rho})}\frac{H^2}{R^2}}.
\end{align}
We introduce a small random perturbation to the initial velocities to initialize the instability. 
Our simulated disk, thus, has the same overall structure as the disks in the simulations by \citet{Nelson2013} and \citet{Manger2018}.
The simulations are conducted with a resolution of $64/H$ (grid cells/pressure scale height) in a spherical coordinate system. In radius, our domain is approximately centered at the midplane water ice line (for this model at \SI{5}{\AU}) and spans $\pm 5 H$ radially, and $\pm 5 H$ vertically. At the boundaries, initial values for density and pressure are constantly set to the initial conditions, while the normal component of velocity is subject to reflecting boundary conditions. In that way, temperature is kept constant at the vertical boundaries.
In azimuth, periodic boundary conditions are employed.
In order to reproduce the radial midplane stratification around the ice line in \autoref{fig:Map}, we chose $h_0=H_0/L_0=0.054$, $\beta_T=-1$, and $\beta_{\rho}=-1.5$ for the first simulation. 
With this definition of $h_0$, our simulation domain is given by $r \in (3.65, 6.35)$ in au, and $\theta \in (1.3008,1.8408)$, where $\theta=\pi/2$ corresponds to the disk midplane.
Since cooling times depend also on the opacity and density of the material, we set $\rho_0=\SI{2.306e-11}{\gram \per \cubic \centi \meter}$, to \rev{approximate} the conditions in the previously modeled disk. The opacities are calculated following \citet{Bell1994}.

\subsection{Simulation Analysis}
To assess the turbulence properties in our numerical experiments, we measure the volume averaged vertical velocities in our simulations, as well as the Reynolds stresses and the growth rate in the linear growth phase of the instability. 

The volume averaging of a measured quantity $\Phi$ is chosen to compensate for the unequal grid cell volume in our spherical grid
\begin{equation}
    \langle \Phi\rangle = \frac{\int_V \Phi \,\mathrm{d}V}{\int_V \mathrm{d}V}=\frac{\sum\limits_{i,j,k}^{I,J,K} \Phi_{i,j,k} \, \delta V_{i,j,k}}{\sum\limits_{i,j,k}^{I,J,K} \, \delta V_{i,j,k}},\label{eq:spatialaverage}
\end{equation}
where $\delta V_{i,j,k}$ refers to the volume of the grid cell with spatial indices $i$ (radial coordinate), $j$ (polar coordinate), and $k$ (azimuthal coordinate, only relevant for three-dimensional simulations). The summation extends over the whole analyzed simulation domain with $I$ cells in the radial direction,  $J$ cells in the vertical direction and $K$ cells in the azimuthal direction. This simple averaging method can also be executed over only the radial sub domain to extract the vertical profile of the analyzed quantity, e.g. the vertical velocity profile.
To calculate Reynolds stresses from our simulation output, we use a similar method as discussed in \citet{Klahr2003} and \citet{Manger2018}, where $\langle\rangle_t$ refers to a time average, and $\langle\rangle_{R,z}$ refers to spatial volume averages as defined above. The spatial distribution of the Reynolds stress of the simulation output with time index $N$ is calculated from time averages, following
\begin{align}
    \alpha_{N,i,j,k} &= \frac{\langle\rho v_r v_{\phi}\rangle_t - \langle v_{\phi}\rangle_t \langle\rho v_r\rangle_t}{\langle P \rangle_t} \\
    &= \frac{\frac{1}{N}\sum\limits_{n=0}^N \rho_{n} \, v_{r,n} \,  v_{\phi,n}-\frac{1}{N^2}\sum\limits_{n=0}^N v_{\phi,n} \cdot \sum\limits_{n=0}^N \rho_{n} \, v_{r,n}}{\frac{1}{N}\sum\limits_{n=0}^N P_{n}}
\end{align}

The spatially averaged Reynolds stress is calculated using \autoref{eq:spatialaverage} with $\Phi=\alpha_{N,i,j,k}$.

For the calulation of the instability's growth rate from our two-dimensional simulations $\Gamma$, we fit the exponential function
\begin{equation}\label{eq:exp}
    \epsilon(\epsilon_0, \Gamma,t, t_0)=\epsilon_0 \exp{(\Gamma(t-t_0))}
\end{equation}
to the linear growth phase of the volume averaged specific kinetic energy
\begin{equation}
    \langle \epsilon \rangle(t) = \left\langle \frac{1}{2} \rho (v_R^2 + v_z^2) \right\rangle_{R,z}, 
\end{equation}
using a non-linear least squares method. 

All numerical evaluations are performed using the standard {\it Python} packages {\it NumPy} \citep{Numpy2011} and {\it SciPy} \citep{2020SciPy}.

\section{Two-dimensional Simulations}
\label{sec:ReferenceSimRelax}

\begin{figure}[ht]
    \centering
    \includegraphics[width=0.45\textwidth]{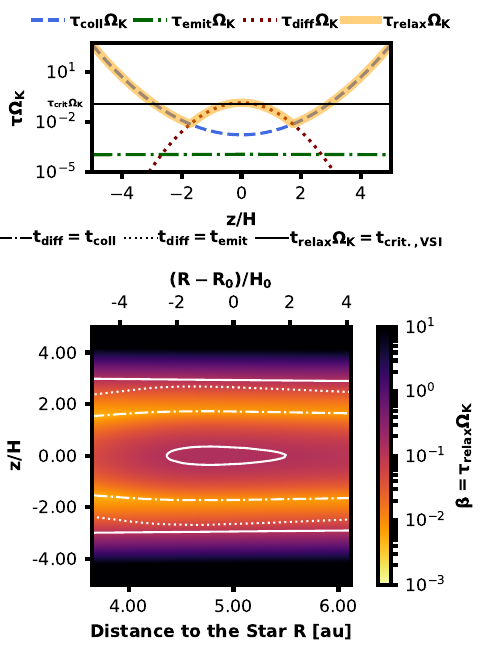}
    \caption{Distribution of the thermal relaxation times, calculated for our simulation domain. In the upper row, the optically thin emission timescale of the dust, the radiative diffusion timescale (for $kH=20$), and the dust-gas collision timescale are shown.
    The total relaxation time (from \autoref{eq:trelax}) is shown in the lower plot. The thick white lines mark the locations at which the rate of thermal relaxation reaches the critical value, above which the VSI is not able to grow efficiently. The very upper layers, as well as the inner, optically thick region are thus not susceptible to the VSI.
    The dash-dotted lines  mark the locations at which the collisional timescale between dust and gas particles becomes as large as the diffusion timescale. At the dotted line, the diffusion timescale becomes equal to the thermal emission timescale, and thus marks the transition from optically thick to optically thin thermal relaxation \rev{for modes with $kH=20$}.}
    \label{fig:RelaxationTimes}
\end{figure}

We perform two-dimensional simulations of local patches in the $r$-$\theta$ plane of stratified PPDs. The relaxation times are calculated in every timestep following the method introduced in the previous sections. For the parameters of the simulations presented and discussed in the respective sections, see \autoref{tab:RelaxParameters}. \autoref{fig:RelaxationTimes} depicts the distribution of thermal relaxation times in a disk with a radiative diffusion wavenumber (i.e. the assumed wavelength), of $kH=20$. The three relevant cooling timescales are shown in the upper row of the figure. It can be seen that the thermal emission timescale is the shortest available timescale everywhere in the disk, and therefore never sets an upper limit on the relaxation timescale of the gas. The large panel depicts the timescale at which the gas is cooled, depending on the location in the simulation, i.e. the maximum timescale of the three timescales shown above. Within $\approx 1$ pressure scale heights of the disk, radiative diffusion limits the cooling time. In the upper layers of the disk, the optical depth decreases strongly, but due to the collisional decoupling of gas and dust particles, cooling is nonetheless slow. 
The VSI can thus only operate within a certain distance from the midplane, and is also suppressed close to the midplane, where radiative diffusion is slow, due to the increasing optical depth in some cases. \rev{An opacity maximum can be found around the center of the simulated region, which is centered at the ice line. At this location, thermal relaxation is accordingly less efficient, and the linear growth of the VSI is inhibited around the central midplane region. Note, however, that smaller VSI modes (larger $kH$), might still be able to grow at these locations, as our relaxation time map in \autoref{fig:RelaxationTimes} only depicts the relaxation times for a fixed $kH=20$.}

\begin{table*}[ht]
\centering
\caption{Parameters of the simulations performed with prescribed thermal relaxation.}
\begin{tabular}{lcccccccc}
\toprule
Section & $M_* [\si{\Msol}]$ & $\beta_T$ & $\beta_{\rho}$ & $R_0 [\si{\AU}]$ & $H_0/R_0$ & $\rho_0 [\si{\gram \per \cubic \centi \meter}]$ & $kH$ & $\frac{H\rev{_0}/R_0}{\Delta \theta}$ \\
\midrule
\midrule
\ref{sec:ReferenceSimRelax} & 1.0 & -1 & -1.5 & 5 & 0.054 & $2.306\times 10^{-11}$ & 20 &  64 \\
\midrule
\ref{sec:WavenumberStudy} & 1.0 & -1 & -1.5 & 5 & 0.054 & $2.306\times 10^{-11}$ & 1 & 64 \\
\ref{sec:WavenumberStudy} & 1.0 & -1 & -1.5 & 5 & 0.054 & $2.306\times 10^{-11}$ & 5 & 64 \\
\ref{sec:WavenumberStudy} & 1.0 & -1 & -1.5 & 5 & 0.054 & $2.306\times 10^{-11}$ & 10 & 64 \\
\ref{sec:WavenumberStudy} & 1.0 & -1 & -1.5 & 5 & 0.054 & $2.306\times 10^{-11}$ & 12 & 64 \\
\ref{sec:WavenumberStudy} & 1.0 & -1 & -1.5 & 5 & 0.054 & $2.306\times 10^{-11}$ & 14 & 64 \\
\ref{sec:WavenumberStudy} & 1.0 & -1 & -1.5 & 5 & 0.054 & $2.306\times 10^{-11}$ & 16 & 64 \\
\ref{sec:WavenumberStudy} & 1.0 & -1 & -1.5 & 5 & 0.054 & $2.306\times 10^{-11}$ & 18 & 64 \\
\ref{sec:WavenumberStudy} & 1.0 & -1 & -1.5 & 5 & 0.054 & $2.306\times 10^{-11}$ & 22 & 64 \\
\ref{sec:WavenumberStudy} & 1.0 & -1 & -1.5 & 5 & 0.054 & $2.306\times 10^{-11}$ & 24 & 64 \\
\ref{sec:WavenumberStudy} & 1.0 & -1 & -1.5 & 5 & 0.054 & $2.306\times 10^{-11}$ & 30 & 64 \\
\ref{sec:WavenumberStudy} & 1.0 & -1 & -1.5 & 5 & 0.054 & $2.306\times 10^{-11}$ & 40 & 64 \\
\midrule
\ref{sec:StratificationStudy} & 1.0 & -0.5 & -2.1 & 5 & 0.054 & $2.306\times 10^{-11}$ & 20 & 64 \\
\ref{sec:StratificationStudy} & 1.0 & -0.6 & -2.1 & 5 & 0.054 & $2.306\times 10^{-11}$ & 20 & 64 \\
\ref{sec:StratificationStudy} & 1.0 & -0.7 & -1.5 & 5 & 0.054 & $2.306\times 10^{-11}$ & 20 & 64 \\
\ref{sec:StratificationStudy} & 1.0 & -0.8 & -1.5 & 5 & 0.054 & $2.306\times 10^{-11}$ & 20 & 64 \\
\ref{sec:StratificationStudy} & 1.0 & -0.9 & -1.5 & 5 & 0.054 & $2.306\times 10^{-11}$ & 20 & 64 \\
\midrule
\ref{sec:3D} & 1.0 & -1 & -1.5 & 5 & 0.054 & $2.306\times 10^{-11}$ & 20 & 32 \\
\bottomrule
\end{tabular}
\label{tab:RelaxParameters}
\end{table*}

\begin{figure*}[ht]
    \centering
    \includegraphics[width=\textwidth]{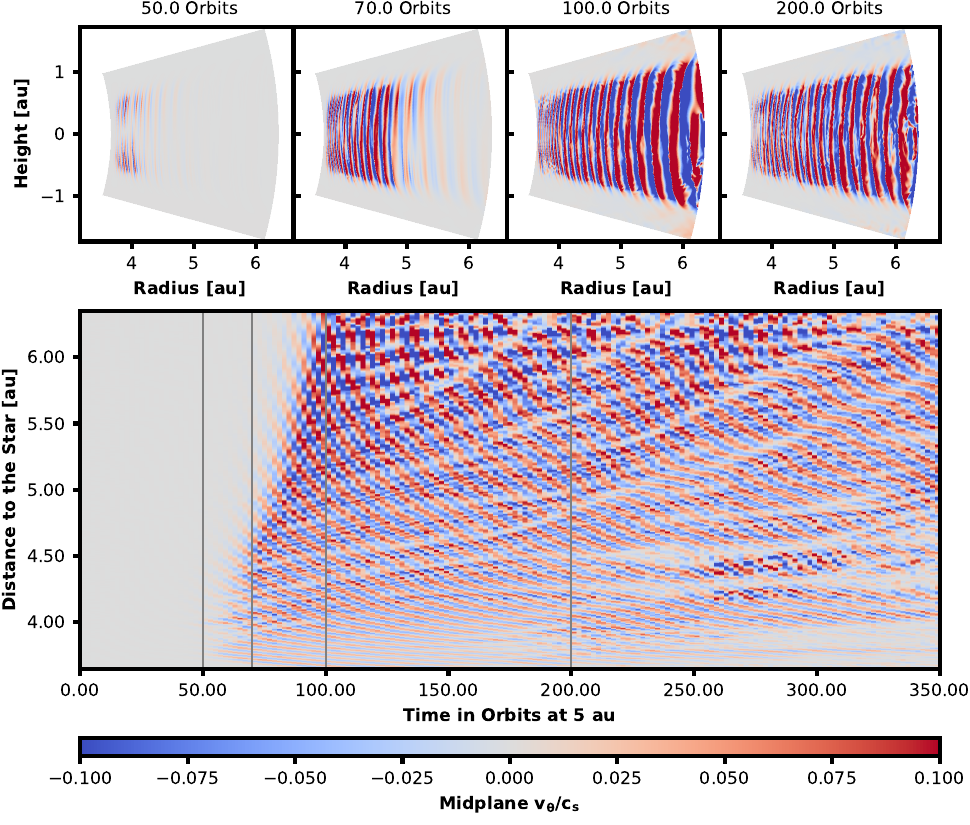}
    \caption{Time evolution of the two-dimensional reference simulation. The upper four panels show snapshots of the polar velocity in simulation at 50, 100, 400 and 900 orbits. In the lower panel, I show the vertically averaged polar velocity as a function of simulation time and distance to the star. The instability first develops antisymmetric flow structures (with respect to the midplane), that evolve into motion that are extend over the full vertical size of the disk. The persistent up and down flowing structures can the be seen to travel inward in the lower panel. As can be seen reflections lead to some unwanted effects at the boundary. We therefore evaluate the simulations only inside of the simulation domain.}
    \label{fig:Timeseries}
\end{figure*}

This effect can clearly be seen in the upper panels of \autoref{fig:Timeseries}, where the time evolution of the polar velocity of the gas in our first two-dimensional simulation is shown.
The VSI first creates zonal flows forming from the top layers of the disk that can still cool sufficiently fast. These nearly vertical flows then progress towards the midplane, where they merge with their counterparts from the opposite disk hemisphere. This evolutionary pattern was also observed by \citet{Nelson2013}, and \citet{Stoll2014}. However, in their simulations with spatially constant cooling times, turbulence started to develop right away at the upper and lower boundary of the simulation domain where the vertical shear is strongest.
In our case, the very upper parts of the disk can not cool sufficiently fast to compensate the entropy differences between the up- and down-welling gas parcels and the background disk. The zonal flows therefore experience repelling buoyancy forces, leading to the suppression of the VSI in the upper layers.
The turbulence created by the VSI in our simulations is thus confined to the efficiently cooling layer and constrained by the inefficient relaxation times in the upper layers, caused by the collisional decoupling of gas and dust particles.

The lower panel of \autoref{fig:Timeseries} shows the time evolution of the midplane value of the $\theta$ component of the velocity, as a function of radius. 
Similar to the numerical experiments by \citet{Stoll2014}, we observe the flow structures to slowly travel inwards, with some minor, unwanted reflection effects at the inner boundary. We therefore decided to only evaluate the gas dynamics and turbulence properties of the inner parts of the disk, between \SIrange{4.5}{5.5}{\AU}. 
Several disruptions in the radial pattern can be seen as light stripes in the time evolution. These phase jumps occur due to the radially varying dominant wavenumber of the VSI and were also observed by \citet{Stoll2014}.

\rev{The radial migration of the VSI pattern is not related to radial mass flux (accretion) in the disk, which is about an order of magnitude slower than the pattern speed. The pattern speed is therefore the phase velocity of the vertical oscillations in the non-linear state of VSI \citep{Nelson2013, Stoll2014}.
Qualitatively one can understand this effect from the quadratic increase in vertical shear with height above the midplane. As a result the VSI driving modes are getting stronger and stronger bend outward with height. This means that any downwards motion at a given height is driven from the stronger shear above and therefore occurs on a shallower angle with respect to the midplane than the corresponding upward motion, which is driven from the more vertical VSI modes from below. In average gas parcels are therefore performing a zig-zag pattern moving less radially outward in their upward motion than what they move radially inward in the down draft. This effect can also be seen in the steepness of the radial gradients of vertical velocity. Streamlines moving down try to get as close as possible to the next inner streamline moving upward.}

\begin{figure*}[ht]
    \centering
    \includegraphics[width=\textwidth]{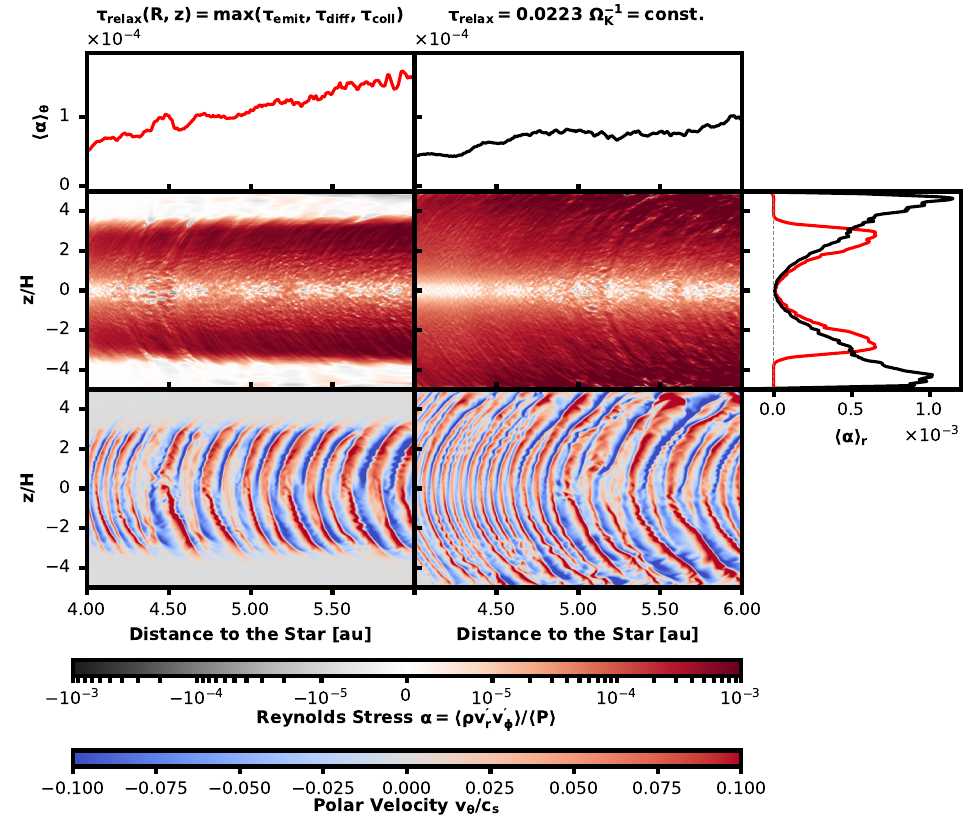}
    \caption{Comparison between a simulation with the refined relaxation time model (left column) and a simulation with a fixed cooling rate (middle column). The upper row depicts the vertically averaged Reynolds stress. The rightmost panel shows the radially averaged Reynolds stresses in the simulation with our new thermal relaxation model (red), and the simulation with constant cooling (black). The middle row shows the spacial distribution of the Reynolds stress. 
    Red color in the figures in the second row indicates outward transport of angular momentum.
    In the lower row, we depict the vertical velocities in the disk after 300 orbital timescales. The up- and down-streaming zonal flows are clearly cut off in the left column, due to the reduced cooling efficiency caused by the collisional decoupling of dust and gas particles in the atmosphere, which is not the case in our simulations with constant cooling rates (middle column).}
    \label{fig:Stress}
\end{figure*}

To visualize the effect of the new relaxation time model further, and to analyze the turbulent properties of the gas, we plot the Reynolds stresses in our two-dimensional simulation in \autoref{fig:Stress}. 
The left-hand side color maps show the stresses induced by the VSI turbulence in our simulation with the new relaxation time mode and the respective polar velocities, with a clear cut-off at $\approx 3$ pressure scale heights above/below the midplane. In contrast, the right-hand side color maps show the stresses and velocities in a simulation with a spatially fixed cooling time, which is fully turbulent up to the vertical simulation boundary. 
In the upper two panels of \autoref{fig:Stress} we show the vertically averaged Reynolds stresses, which can be seen to slightly increase with distance to the star. In both simulations, $\alpha\approx \SI{1e-4}{}$ is reached, in accordance with earlier studies \citep{Stoll2014}. In the right panel, we plot the radial average of the stresses, depicting the simulation with the new cooling model in red, and the simulation with the spatially constant cooling in black. The vertical cut-off can again be seen to occur at $\approx 3$ pressure scale heights in out new model, with a maximal stress of $\alpha \approx \SI{6e-4}{}$.

Two-dimensional axisymmetric flow structures are typically observed to induce negative Reynolds stresses in numerical experiments, e.g. in simulations of convectively unstable disks \citep[see][and references therein]{Klahr2007}. The VSI, however, introduced a mostly vertical transport of angular momentum \citep{Manger2018}. With height above the midplane, the flow's direction becomes more and more radial due to the parabolic shape of the surfaces of constant angular momentum. This can also be seen in the lower panels of \autoref{fig:Stress}, where we depict the vertical velocities after 300 orbits of evolution. In that way, angular momentum is effectively transported away from the midplane, and radially outwards in the atmosphere. The result is a vertically increasing, positive Reynolds stress, even for the axisymmetric VSI structures in our two-dimensional simulations.

\subsection{Dependency on the Diffusion Timescale/Length Scale}
\label{sec:WavenumberStudy}
The thermal relaxation times in the inner parts of our simulations, where both collisions and optically thin emission are very effective for the tested parameters, are limited by the speed of radiative diffusion.
As any diffusive process, radiative diffusion happens on a timescale that depends on the spatial extent of the underlying perturbation in the diffusing quantity -- in our case pressure, or temperature. To calculate the cooling time in the optically thick parts of the disk according to \autoref{eq:thick}, we thus have to decide on what length scale we chose to \rev{approximate} the diffusive cooling that a typical VSI flow structure would undergo. Simulations of the optically thick parts of the disk, performed by us with flux-limited radiative diffusion (FLD) \citep{Levermore1981}, have shown that the arising VSI zonal flows have approximate wavenumbers of $kH\sim 20$, for the given parameters and at the studied location in the disk (see Appendix). 
This value of course changes with location, as it depends on the optical depth and the rate of vertical shear. For this first study, we, however, keep it a constant in the whole simulation domain. 
In order to investigate in how far the choice of this radiative diffusion length scale influences the outcome of our simulations, we performed two-dimensional simulations with the same parameters as in the previously presented test case, and for different diffusion wavenumbers of $k=\SIrange{1}{40}{H^{-1}}$. 

Note, that a wavenumber of $kH=1$ corresponds to a physical size of $\lambda = 2\pi/k = 2\pi H \approx \SI{1.7}{\AU}$ at \SI{5}{\AU} distance to the star -- a rather extreme, and unrealistic size for a zonal flow caused by a comparably weak instability like the VSI at this location. In contrast, a wavenumber of $kH=40$, corresponds to a size of $\lambda \approx 0.16 H \approx \SI{0.0432}{\AU}$, which is only resolved by 10 grid cells in our simulation (i.e. 5 cells per up- or down-welling stream), but was observed by us in our FLD simulations for the highest tested gas density of $\rho_0=\SI{1e-10}{\gram \per \cubic \centi \meter}$.
\citet{Lin2015}, however, suggest typical sizes of $kH\sim 10$, as they find much smaller structures to be damped by viscosity. 

Since our simulations lack any physical implementation of the underlying diffusive energy transfer, we do not introduce an additional physical scale, as it would be the case for simulations that explicitly treat radiative diffusion. This means structures of any size are cooled at the same rate in our simulations, which would be nonphysical if an unrealistic diffusion parameter $k$ would be chosen. 
We therefore test in how far the choice of $k$ influences the finally achieved gas velocities and Reynolds stresses to see if our $\beta$-cooling model with $kH=20$ leads to results as close as possible to a simulation with flux-limited diffusion in the disk's interior.

\begin{figure}[ht]
    \centering
    \includegraphics[width=0.45\textwidth]{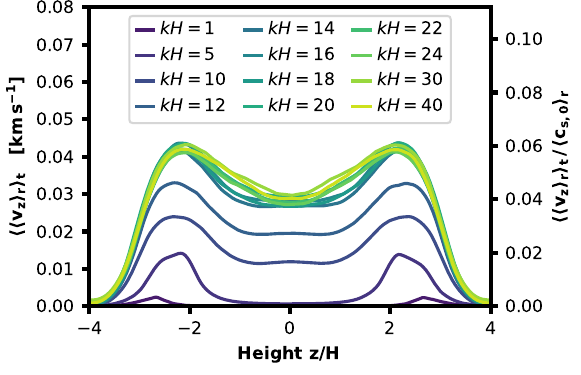}
    \caption{Radially and time-averaged vertical velocities of simulations with different radiative diffusion wavenumber.}
    \label{fig:vzk}
\end{figure}

\autoref{fig:vzk} depicts the radially averaged vertical profile of the vertical velocity for the simulations with different wavenumbers in \si{\kilo \meter \per \second}.
In general, higher applied wavenumbers correspond to smaller structure and, thus, more efficient cooling.
We only find a significant influence of the wavenumber on the vertical gas velocities for wavenumbers smaller than $kH=14$, as can be seen in \autoref{fig:vzk}. The maximally reached vertical velocities for higher wavenumbers than $kH=14$, are $v_{z, \text{max}}\approx \SI{0.04}{\kilo \meter \per \second}$. Velocities are are generally lower for lower wavenumbers, which correspond to slower cooling. 

\begin{figure*}[ht]
    \centering
    \includegraphics[width=\textwidth]{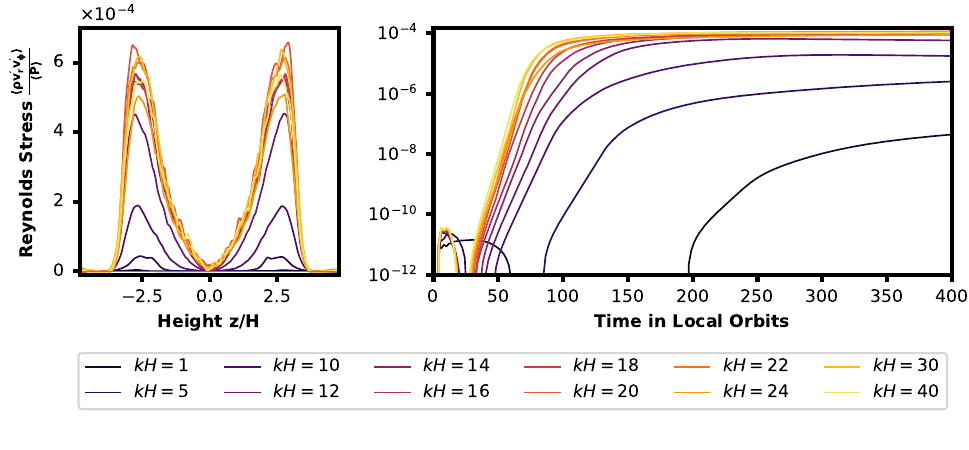}
    \caption{Dependency of the Reynolds Stress on the diffusion time control parameter $kH$. A larger $k$ corresponds to a faster diffusion timescale. The Reynolds Stresses are in the order of $10^{-4}-10^{-5}$, increasing with faster cooling.}
    \label{fig:Alphak}
\end{figure*}

The same is true for the Reynolds stresses achieved after saturation of the turbulence, as depicted in \autoref{fig:Alphak}.
Similar to the vertical velocity profile, we measure a double-peaked profile for the Reynolds stresses caused by slowly cooling atmospheric layer. Simulations with a diffusion wavenumber $kH>14$, reach average stresses of $\alpha \sim \SI{1e-5}{}$, with maximal values of $\alpha\approx \SIrange{5e-4}{6.5e-4}{}$ at $\approx 2.5$ pressure scale heights distance from the midplane. 
All simulations shown here have an upper cut-off of the VSI turbulence due to the transition to poor coupling between dust and gas particles at $\SIrange{2}{4}{H}$.

\begin{figure*}[ht]
    \centering
    \includegraphics[width=\textwidth]{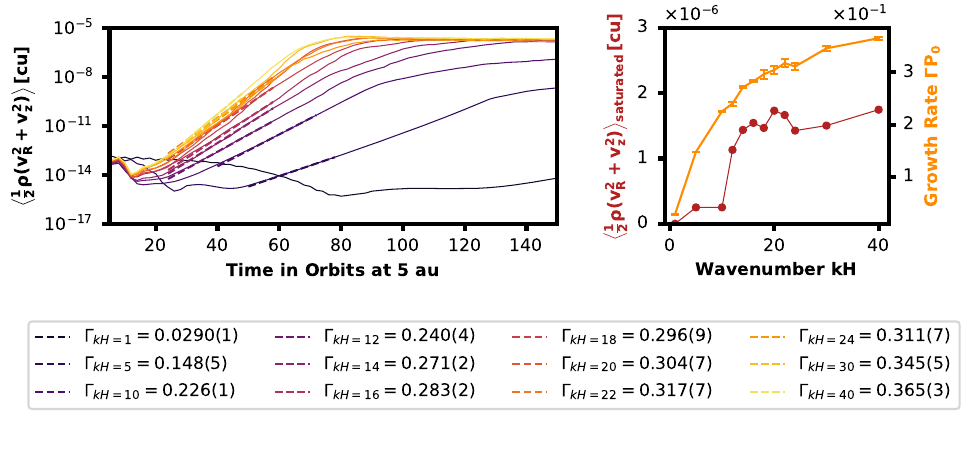}
    \caption{Determination of the growth rates of the instability within an interval of \SI{0.5}{\AU}, centered in the domain. The larger the wavenumber, the faster the cooling of the VSI modes. Thus, growth rates increase with wavenumber. At $kH\sim30$, cooling becomes so efficient that every perturbation is essentially isothermal. The dependency of the VSI's growth rate on the wavenumber therefore levels of, as even faster cooling makes no difference to the isothermal motions.}
    \label{fig:Growthratek}
\end{figure*}

\subsubsection{Growth Rates}
We also measure the growth rates of the VSI in dependence of the diffusion wavenumber.
In \autoref{fig:Growthratek}, we show the time evolution of the mean specific kinetic energy $\epsilon = 0.5\rho (v_r^2+v_\theta ^2)$ of the gas for the different simulations runs. 
Time in our simulation is given in units of the local orbital period $P_0=\sqrt{4\pi^2 R_0^3 / G M_*}$ at \SI{5}{\AU}, which means the growth rates determined by us are are given in units of $P_0^{-1}$. 
For our reference simulation with $kH=20$, we find $\Gamma P_0= 0.304(7)$. Increasing the wavenumbers leads to faster growing VSI modes, because of the more efficient cooling in the interior of the disk. For $kH=40$, we get $\Gamma P_0 =0.365(3)$ and for the slowest cooling, i.e. $kH=1$, we find $\Gamma P_0=0.0290(1)$. We observe the growth rates dependency on the diffusion wavenumber to slightly level off for large wavenumbers. The reason for this is that all perturbations are essentially isothermal for such fast cooling. The same is true for the final value of the mean kinetic energy, which is strongly rising up to $kH=20$ and then levels of for larger wavenumbers.

We, thus, find that the cooling time is an essential parameter, that has great impact on the velocities of the VSI flow structures, the growth rates of the instability and the Reynolds stresses.
However, for wavenumbers in the order of $kH \geq 14$, the resulting resulting turbulence reaches similar maximum velocities and stresses. If the cooling is chosen to be less efficient ($kH\leq 14$), growth rates and turbulent velocities are smaller due to the stronger influence of the repelling buoyancy forces on the VSI modes.

\subsection{Dependency on the Radial Stratification}
\label{sec:StratificationStudy}
The VSI is crucially dependent on the rate of vertical shear that exists in PPDs. This vertical shear, in turn, is strongly influenced by the radial stratification in temperature, as becomes evident from \autoref{eq:RHE}.
Observations and numerical modeling of PPDs shows that a variety of radial temperature gradients can be present \citep{Andrews2007, Pfeil2019}. From $\beta_T=-0.5$, for a passively irradiated disk, to $\beta_T=-1$ for a disk that is strongly heated in the midplane by viscous dissipation. 
It is evident from linear stability analysis of a vertically isothermal disk, that the growth of the VSI is directly proportional to the radial gradient in temperature  \citep{Urpin1998, Nelson2013, Lin2015}

\begin{equation}
    \Gamma P_0 \propto |\beta_T| \frac{H}{R}.
\end{equation}

To test this result for the VSI in our simulations with more realistic thermal relaxation, we perform simulations for $\beta_T=-1,\, -0.9,\, -0.8,\, -0.7,\, -0.6,\, -0.5$.
The radial density gradient was shown to have no influence on the growth rate of the VSI by other authors \citep{Nelson2013, Manger2018, Manger2020}, and we therefore do not perform a systematic parameter study for this disk property. We thus set $\beta_{\rho}=-2.1$, for the temperature gradients that resemble passively irradiated disk ($\beta_T=-0.6, \, -0.5$), and $\beta_\rho=-1.5$ for the temperature slopes that resemble the structure of viscously heated disks ($\beta_T=-1,\, -0.9,\, -0.8,\, -0.7$).

\begin{figure}[ht]
    \centering
    \includegraphics[width=0.45\textwidth]{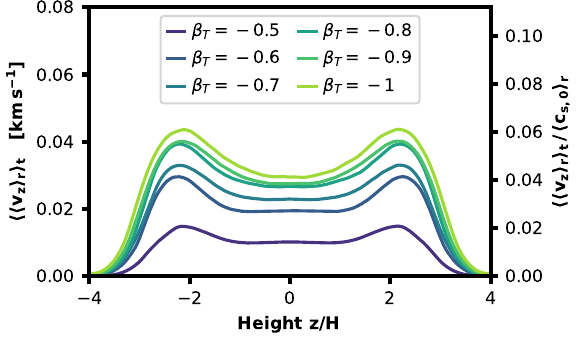}
    \caption{Vertical distribution of vertical velocities for different stratifications. The vertical velocity was averaged over the different radii and a suitable time interval to account for the different growth rates.}
    \label{fig:vzstrat}
\end{figure}

\begin{figure*}[ht]
    \centering
    \includegraphics[width=\textwidth]{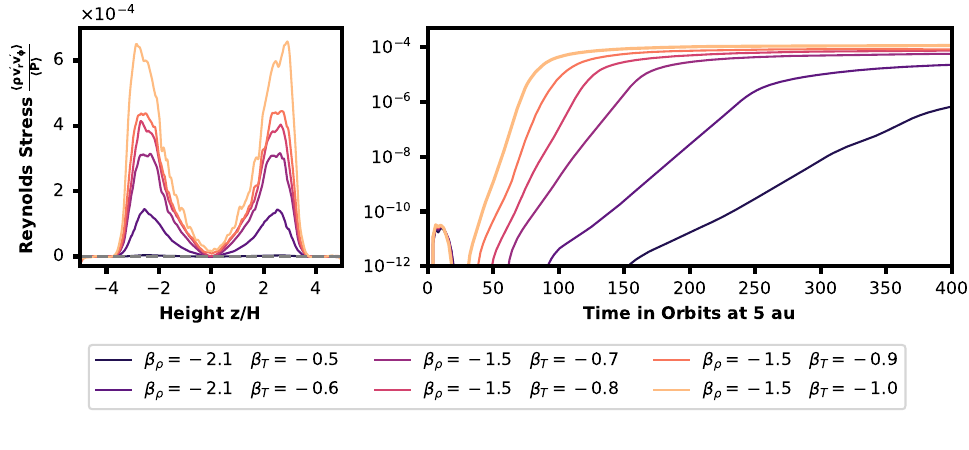}
    \caption{Dependency of the Reynolds Stress on the stratification of the disk. Here we cover the relevant range of the temperature gradient for PPDs, from $-0.5$ (passively irradiated disk), to $-1$ (very strong viscous heating). A steeper gradient in temperature leads to a higher final Reynolds Stress. The steeper stratified disks have Reynolds Stresses in the order of $10^{-4}-10^{-5}$, which decrease slightly if the temperature gradient is chosen to be more shallow.}
    \label{fig:AlphaStrat}
\end{figure*}
\begin{figure*}[ht]
    \centering
    \includegraphics[width=\textwidth]{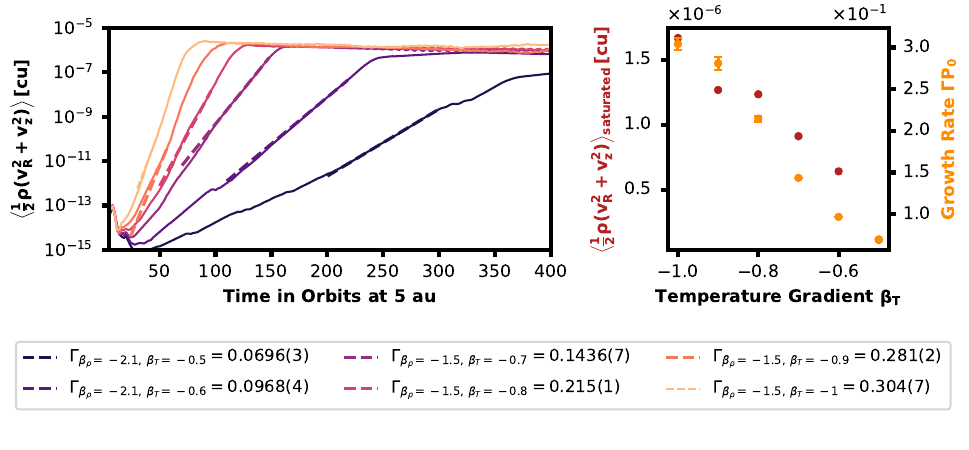}
    \caption{Growth rates of the VSI for different stratification. The steeper the temperature gradient of the disk, the higher the growth rate. Here, we recover the predicted linear dependency of the growth rate on the temperature gradient.}
    \label{fig:GrowthRateStrat}
\end{figure*}

As expected, we find generally higher vertical velocities for the simulations with steeper radial temperature slope, as shown in \autoref{fig:vzstrat}. For $\beta_T=-1,\, -0.9, \, -0.8$, maximal velocities reach $\approx \SI{0.04}{\kilo \meter \per \second}$ at $\approx 2.2$ pressure scale heights above/below the midplane. A radial gradient of $\beta_T=-0.7$ or $\beta_T=-0.6$, resulted in $v_{z,\text{max}}\approx \SI{0.03}{\kilo \meter \per \second}$ and for the shallowest profile tested, $\beta_T=-0.5$, we find $v_{z,\text{max}}\approx \SI{0.015}{\kilo \meter \per \second}$.

Similarly, Reynolds stresses are generally higher for steeper temperature stratification, as depicted in \autoref{fig:AlphaStrat}. The volume averaged stresses are generally of order $\sim \SIrange{1e-6}{1e-4}{}$, increasing for a steeper stratifications.
The maximum stress levels are reached at a height of $\approx 2.5$ pressure scale heights from the midplane and range from $\alpha \approx 10^{-5}$, for $\beta_T=-0.5$ to values of the order $\sim 10^{-4}$ for steeper stratification.

\subsubsection{Growth Rates}
In \autoref{fig:GrowthRateStrat}, we plot the time evolution of the averaged specific kinetic energy in our simulations with different radial temperature stratification.
To determine the instability's growth rate, we fit the exponential function \autoref{eq:exp} to the growth phase of the VSI. For the shallowest radial profile ($\beta_T=-0.5$), we get a growth rate of $\Gamma P_0= 0.0696(3)$. The rates are linearly increasing with steeper temperature gradients up to $\Gamma P_0 =0.304(7)$ for the steepest gradient of $\beta_T=-1$. 
A similar behavior can be seen for the final value of the mean specific kinetic energy, which is larger for steeper gradients in temperature.
We can, therefore, confirm the linear dependency of the VSI's growth rate on the radial temperature gradient in our simulations.

\section{Three-dimensional Simulation}
\label{sec:3D}
In this section, we focus on the evolution of the VSI in three-dimensional simulations, including the formation of long-lived anticyclonic vortices.
Zonal flows and anticyclonic vortices play an important role in planet formation because they can act as dust traps due their pressure structure. 
Here, we show that they can form in our PPD simulation with realistically prescribed thermal relaxation rates. 
The simulation is carried out with a resolution of $32/H$ in all three dimensions and covers an azimuthal angle of \SI{90}{\degree}. 
Since we aim to study structure formation due to the VSI, the quantity of midplane vorticity is of special interest. To better visualize anticyclonic vorticity perturbations, we normalize the vorticity by the background profile of the disk itself, i.e.
\begin{equation}
\omega_z' = \frac{2\omega_z}{\Omega_{\text{K}}}= \frac{2(\vec{\nabla}\times \vec{v})_z}{\Omega_{\text{K}}}.
\end{equation}
Thus, every value of $\omega_z'$ below 1 in our simulations corresponds to an anticyclonic flow, i.e. a structure counter-rotating relative to the disk's rotation. 

\begin{figure}[ht]
    \centering
    \includegraphics[width=0.45\textwidth]{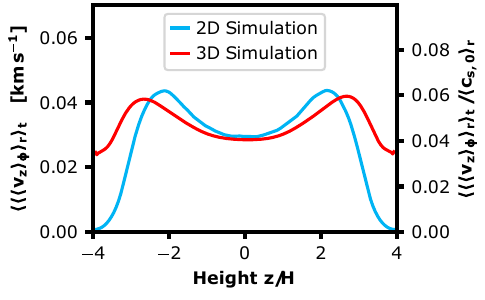}
    \caption{Comparison of the spatially and temporally averaged vertical velocities of the two- and three-dimensional simulations. The two-dimensional simulation domain extends $\pm5$ pressure scale heights vertically, the three-dimensional simulation domain extends $\pm 4$ pressure scale heights vertically.}
    \label{fig:vZCompare}
\end{figure}

The structure of the simulated disk is the same as in our two-dimensional reference simulation with a unit density of \SI{2.306e-11}{\gram \per \cubic \centi \meter}. The simulation domain is radially centered at \SI{5}{\AU}, with a disk aspect ratio of $H/R=0.054$. The stratification follows a radial power law with $\beta_T=-1$ and $\beta_{\rho}=-1.5$. 
The simulation domain presented here, has a vertical size of $\pm 4$ pressure scale heights and spans $\pm 7$ scale heights radially, centered at \SI{5}{\AU}.
In a few tens of orbits, the VSI develops axisymmetric flow structures that grow in intensity with time. The instability starts to grow at a height of $\approx 3$ pressure scale heights, where thermal relaxation times and the rate of vertical shear are most favorable.
This first growth phase is, thus, very similar to our two-dimensional, axisymmetric simulations. The average vertical velocities of the up- and down-flowing streams is also comparable to the results of our two-dimensional simulations. We plot the vertical velocity profile in \autoref{fig:vZCompare}, where both the two-dimensional and the three-dimensional simulation with similar parameters, are shown to reach maximal vertical velocities of $\sim \SI{0.04}{\kilo \meter \per \second}$. We note that the reduced cooling time in the upper atmosphere of the disk also causes a suppression of the VSI in the upper layers of our three-dimensional simulation. However, in three dimensions, non-axisymmetric flow structure, like spiral density waves etc. can form, causing an additional level of atmospheric turbulence.

\begin{figure}[ht]
    \centering
    \includegraphics[width=0.45\textwidth]{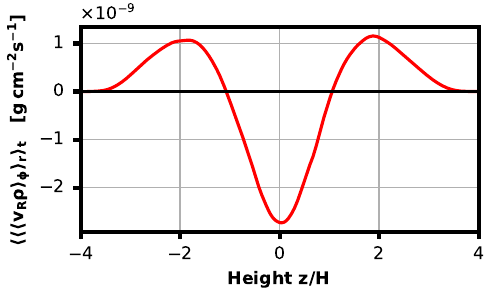}
    \caption{Spatially and temporally averaged radial mass flow in our three-dimensional simulation. In the midplane, mass is flowing inwards, while some mass is flowing outwards in the upper layers, transporting away angular momentum. The average mass flow in our three-dimensional simulation is \SI{-3.426e-11}{\gram \per \square \centi \meter \per \second}, which results in a net mass accretion rate of $\dot{M}=-2\pi R \Sigma v_R = \SI{1.828e-08}{\Msol \per \years}$.}
    \label{fig:Accretion}
\end{figure}

We also measure the average radial mass flux in \autoref{fig:Accretion}. Similar to the results by \citet{Manger2018}, we encounter inward flux in and around the midplane, and outward flux in the upper atmosphere. The net mass flux is directed towards the star, and can be translated into a mass accretion rate of $\dot{M}=-2\pi R \Sigma v_R = \SI{1.828e-08}{\Msol \per \years}$. Thus, by transporting angular momentum mostly vertically upwards and then outwards, the VSI enables inward mass accretion in the midplane, despite very low Reynolds stresses there. 
Viscous accretion disk theory predicts a mass accretion rate of $\dot{M}= 3 \pi \Sigma \alpha v_{\text{K}} R (H/R)^2$ \citep{LyndenBell1974, Pringle1981}. For a typical $\alpha=10^{-4}$, as measured in our three-dimensional simulation, we obtain $\dot{M}=\SI{1.741e-08}{\Msol \per \years}$, which is in very good agreement to the measured value. Similar to \citet{Manger2018}, we find the angular momentum transport caused by the VSI to create mass accretion rates that agree well with the values predicted by classic accretion disk theory, despite the more complex three-dimensional distribution of Reynolds stresses.

\begin{figure*}[ht]
    \centering
    \includegraphics[width=\textwidth]{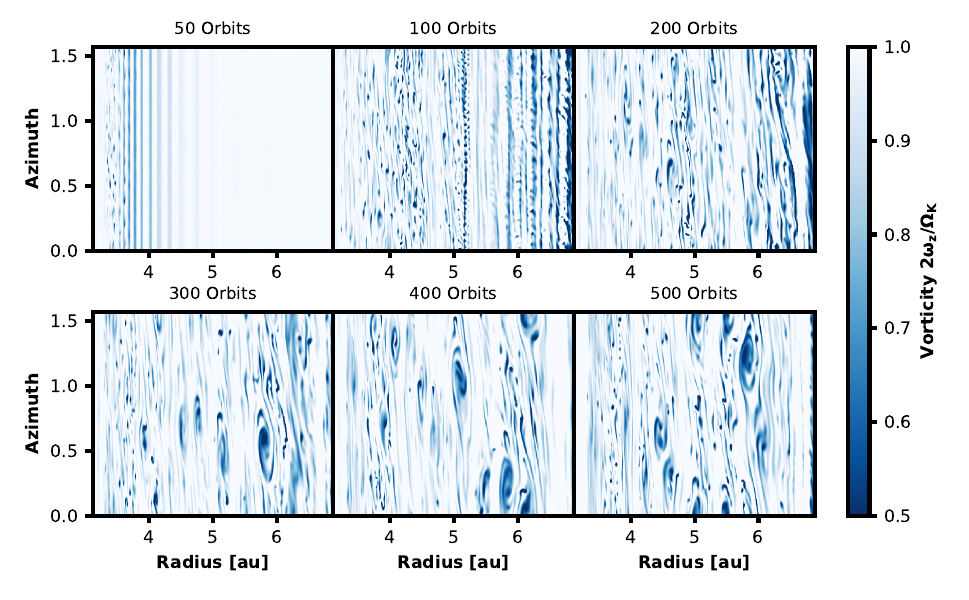}
    \caption{Time evolution of the three dimensional PLUTO simulation.}
    \label{fig:3DTimeEvolution}
\end{figure*}

\autoref{fig:3DTimeEvolution} depicts the time evolution of the midplane vorticity. The VSI first forms flow structures that \rev{show up as an axisymmetric pattern in vorticity. Once the vorticity perturbation violates the local Rayleigh criterion $\omega_z \leq 0$ \citep{Manger2018,Latter2018} small vortices do form from Kelvin-Helmholtz instability (KHI). This first happens in the inner rings, as the development and VSI and KHI is the fastest there (3.5 au at t = 50 orbits) and the effect propagates radially outwards (\SI{5.2}{\AU} at 100 orbits).
The large scale vortices that eventually appear are then a mix of the mergers of small vortices, but also a Rossby Wave Instability (RWI) which can be shown to be triggered for large enough azimuthal extent.
\citet{Manger2018} show that the axisymmetric extrema in vorticity can also cause the RWI, which results in a local break-up of the \rev{axisymmetric VSI} flows. This causes the formation of small anticyclonic vortices which merge and ultimately form large-scale structures. We observe a similar evolutionary pattern as \citet{Richard2016} and \citet{Manger2018}, and find the first long-lived vortices to emerge after 100-250 orbits. }

\begin{figure*}[ht]
    \centering
    \includegraphics[width=\textwidth]{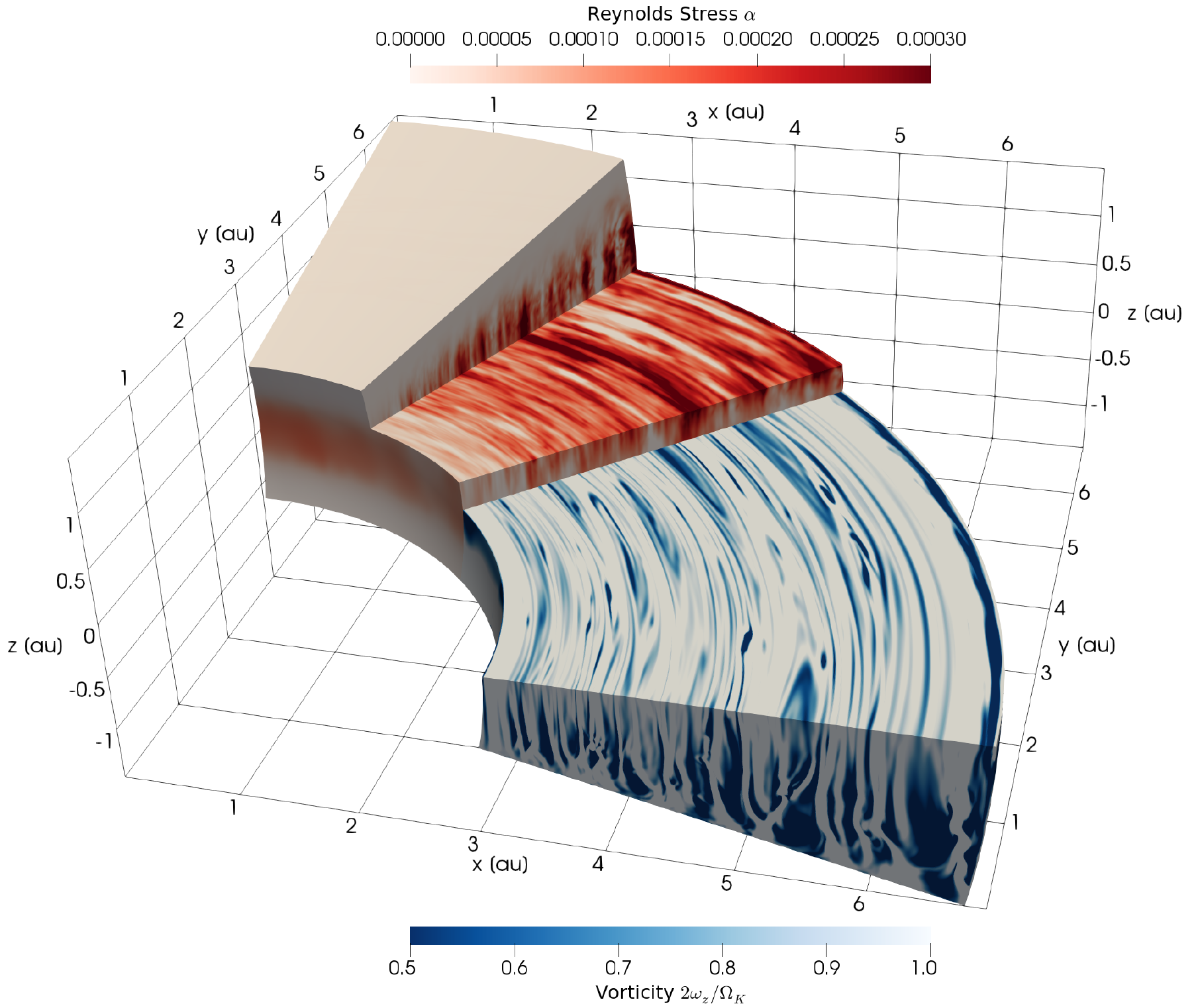}
      \caption{Cut through the three-dimensional simulation after 600 orbits of evolution. The Reynolds stress is strongly increasing with height above the midplane, until the thermal relaxation times becomes larger due to the collisional decoupling of dust and gas particles. Close to the midplane, stresses fall even below $5\times 10^{-5}$. However, VSI turbulence leads to the formation of long-lived vortical structures, even in the weakly turbulent midplane. These structures are visible as blue patches in the cut of the disk and extend over $\pm 3 H$ above/below the midplane until they merge with the turbulent background.}
    \label{fig:3DSnapshot}
\end{figure*}

After 600 orbits, the largest structures have sizes of up to $\sim 10$ pressure scale heights in the azimuth and about $\sim 1$ pressure scale height radially. \rev{These structures can be seen in the three-dimensional depiction of the flow's vorticity in \autoref{fig:3DSnapshot}, where a large anticyclonic vortex appears also in the front cut through the disk. It can be seen that the vortex extends up to $\sim 2$ pressure scale heights below the midplane, deforming the VSI flow structures also in the upper layers. In red, we also show the three-dimensional distribution of the Reynolds stress, which first increases with distance to the midplane, until the inefficient dust-to-gas coupling in the upper atmosphere inhibits the VSI's growth at $\sim 3$ scale heights.}

Here, we also plot the Reynolds stress, caused by the VSI and the non-axisymmetric flow structures. We measure values in the order of $10^{-4}$, similar to previous studies of the VSI and our two-dimensional simulations.

\begin{figure*}[ht]
    \centering
    \includegraphics[width=\textwidth]{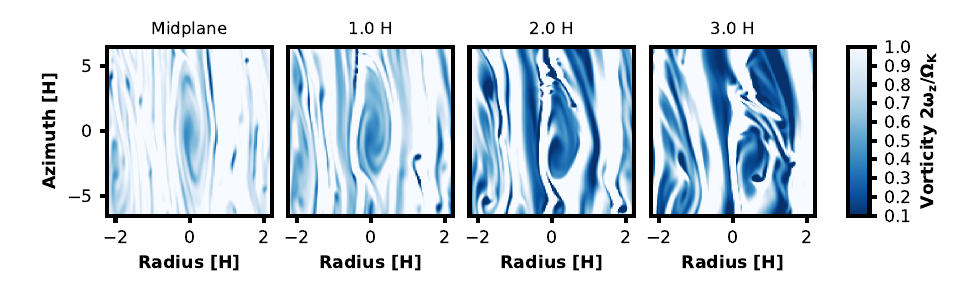}
    \caption{Close-up view of the central vortex at different heights above the midplane after 600 orbits of evolution.
The structure is vertically extended, with only little variety in radial and azimuthal
structure. At 3 pressure scale heights, the vortex blends in with the very turbulent background
flow. These flow features are thus extended over almost $\pm$3 pressure scale heights vertically.}
    \label{fig:3DHeights}
\end{figure*}

\autoref{fig:3DHeights} shows the vertical structure of of the large central vortex after 600 orbits of evolution. Similar to the vortices discovered by \citet{Manger2018}, it has an azimuthally elongated structure, spanning $\sim 10$ scale heights in $\phi$, and $\sim 1$ scale height in radius. The observed vortex is only half as large as the vortices observed by \citet{Manger2018} in their global simulations. The reason for this might be our azimuthally smaller simulation domain. Global simulations with our cooling time prescription should be conducted in the future to study whether a larger simulation domain leads to larger anticyclones, as shown in \citet{Manger2018} for a fixed cooling rate.
The vortex in \autoref{fig:3DHeights} extends over three scale heights above the midplane until it merges with the turbulent background state in the right panel of the figure. 

\begin{figure*}[ht]
\centering
	\includegraphics[width=\textwidth]{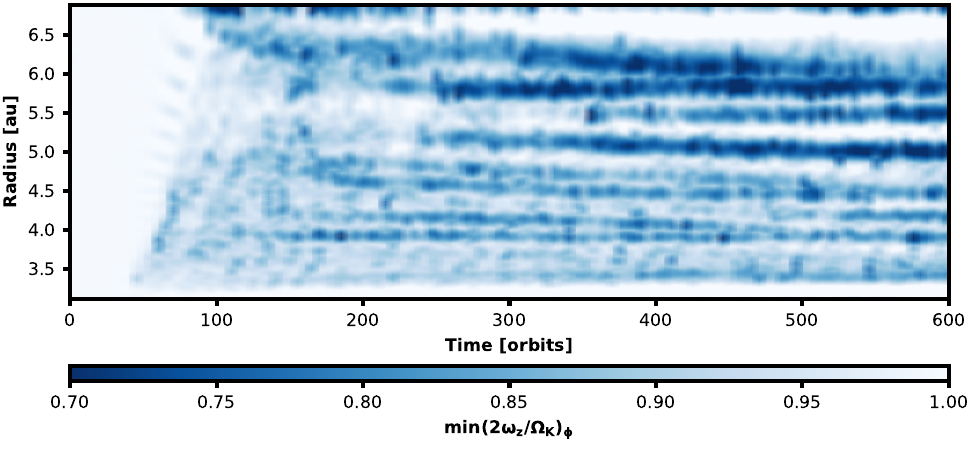}
	\caption{Evolutionary tracks of the azimuthal minima in midplane vorticity in our three-dimensional simulation. Similar to the method presented in \citet{Manger2018},  vorticity has been smoothed out with a Gaussian filter to enable the determination of the large vortices' radial location, without small scale noise. Each continuous blue track corresponds to a long-lived anticyclonic vortex. On the simulated timescale, only minor radial migration can be observed for the central vortex.}
	\label{fig:3DLifetime}
\end{figure*}

In order to study the lifetime of the vortices, we smooth out the vorticity field for all given timesteps by applying a Gaussian filter of width $\sigma=10$ cells, to get rid of small scale fluctuations.
Then, the minimum of the smoothed vorticity field is found over the azimuth at each radius. The result is a time evolution of the local radial minimum in vorticity, shown in \autoref{fig:3DLifetime}, that shows how vortices form, migrate, merge, and how long they survive in the disk. The first long-lived structure in the inner part of the simulation forms at $\approx \SI{4}{\AU}$ after $\approx$ 100 orbits of evolution. This vortex can be seen to merge with a close neighbor at $\approx$ 480 orbits and it is still present after 700 orbits, meaning it survived $\approx 600$ orbits. The vortex pair slightly outside the innermost pair can also be seen to migrate inwards for over 100 orbits.
\rev{Migration has, however, only a minor impact on the overall picture, and seems to be relevant only for adjacent vortex pairs which undergo merging. The reason for the little amount of migration might be the constantly created surface density perturbations due to the VSI. \citet{Meheut2012} also observed very little to no migration at all. The vortices found in our simulations are highly elliptic, with aspect ratios at $\chi \gtrsim 8$, which might also explain their slow migration, as discussed in \citet{Richard2013}.}
Due to their longevity, the observed vortices could act as stable and effective dust traps. 

\begin{figure*}[ht]
    \centering
    \includegraphics[width=\textwidth]{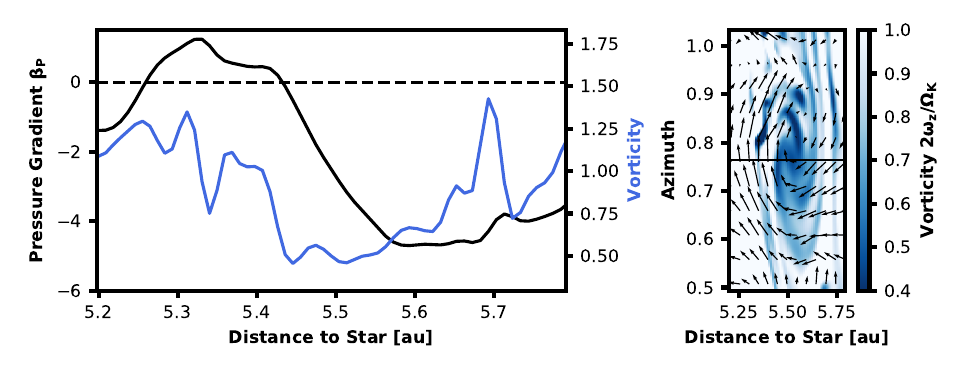}
    \caption{Pressure and vorticity structure of the large central vortex after 720 orbits. The vortex is
visible as a minimum in vorticity and has caused a pressure maximum, which could act as
a dust trap.}
    \label{fig:Vortex_Structure}
\end{figure*}

To investigate this possibility further, we are studying the radial pressure structure induced by the large central vortex. Radially migrating dust grains accumulate in pressure maxima, which makes them potential sites for direct gravitational collapse of the dust clouds or for the triggering of Streaming Instability. The left panel of \autoref{fig:Vortex_Structure} depicts the radial pressure and vorticity profile through the center of the large vortex in \autoref{fig:3DHeights} after 720 orbits. The anticyclone, visible as a minimum in vorticity, has formed a clear maximum in the gas pressure in the midplane of the disk.

Our simulation shows, that VSI turbulence is able to create long-lived anticyclonic vortices with a radial pressure maximum, even in non-isothermal PPDs.

\section{Discussion}
\label{sec:Discussion}
Our studies of the VSI with more realistic thermal relaxation have shown that the instability can operate also in the inner parts of PPDs (around \SI{5}{\AU}), lead to mass accretion in the disk midplane, and form long-lived vortices. 
The achieved Reynolds stresses in the active layer of the disk are consistent with previous studies of the VSI with simpler cooling prescriptions \citep{Nelson2013}, flux limited diffusion \citep{Stoll2014}, or in the outer parts of PPDs \citep{Flock2017, Flock2020, Manger2020}.
In contrast to these studies, we incorporate a simple prescription for the thermal coupling of dust and gas particles, based on the work of \citet{Malygin2017}. The collisional decoupling of dust and gas particles is shown to introduce strong damping of the VSI modes in the upper atmospheres of cold regions of PPDs, due to the reduced rate of thermal relaxation. 
For this, we assume a constant dust-to-gas ratio, as well as dust grains of only one size (\si{\micro \meter} sized). These are radically simplified conditions. In real PPDs, dust populations evolve dynamically and the turbulence of the gas has profound influence on this evolution \citep{Voelk1980, Ormel2007, Birnstiel2009, Johansen2014, Ishihara2018, Gerbig2020, Klahr2020}. Dust grains coagulate, fragment, and -- probably most importantly for this study -- sediment towards the midplane. 
Future work would benefit from a self-consistent treatment of dust evolution. A combination of the hydrodynamic gas evolution with the dust evolution could be used to realistically model the dust opacities, as well as the collisional coupling of the gas and dust species. Both are necessary for a self-consistent simulation of the gas' thermal relaxation, which is essential for the VSI's evolution, as shown in this study.

Furthermore, we have assumed the thermal relaxation of the gas in the upper atmosphere of our simulation to be completely determined by the dust grains' emission. This is only valid, if the respective layers are cold enough, such that the gas' opacities are much smaller the those of the dust. Simulations covering parts of PPDs closer to the central star, or starlight heated upper layers, must consider the thermal timescale of the gas itself in their treatment of thermal relaxation. Our results are therefore only applicable to the outer, cooler parts of PPDs around or beyond the water ice line ($R\gtrsim \SIrange{4}{5}{\AU}$).
Note, that if the dust scale height would be significantly smaller than the gas scale height, the upper layers of PPDs would become depleted of dust.
Consequently, dust-gas collisions would be extremely rare in the upper atmosphere. In this case, thermal relaxation could either become extremely inefficient, if the temperatures are low, or would be dominated by the gas' emissions.

Future studies that aim to incorporate a realistic heating and cooling model, should also include stellar irradiation realistically, like e.g. \citet{Flock2017, Flock2020, Fuksman2020}, in combination with the consideration of thermal dust-gas coupling.

In the interior parts of the disk, close to the midplane, we introduce the radiative diffusion timescale. 
In a simulation incorporating flux-limited radiative diffusion, this process would introduce an upper limit for the size of the emerging flow structure. The reason for this is that a spatially small perturbation in temperature can thermally relax much faster in a diffusive manner, than a large perturbation. In our simulations, we can not simulate this effect, but only regulate the cooling times in the diffusion dominated part of the disk, by pre-setting the diffusion wavelength, which is an input parameter for our model. We have investigated the influence of this length scale on the outcome of our simulations and found values of $kH\gtrsim 20$ to lead to very similar turbulent velocities and Reynolds stresses. 
It would be beneficial for future studies to introduce a self-consistent treatment of heating and cooling. Ideally, a self-consistent stratification in temperature should be achieved, under both the influence of stellar irradiation and viscous heating, in contrast to our very simplified vertically isothermal structure. For this, a more realistic model for both Planck and Rosseland opacities has to be used.

For our simulation of vortex formation, we relied on an azimuthal domain size of \SI{90}{\degree}, which was sufficient to form large scale, long-lived vortices. However, \citet{Manger2018} have clearly shown that a larger domain size has a big impact on the forming structures. Larger structures are generally favored in a larger simulation domain. Our work could, thus, be expanded to a global study of vortex formation with realistic heating and cooling in the future. 

\section{Conclusions and Outlook}
\label{sec:Conclusion}
For the first time, we have conducted two- and three-dimensional simulations of the VSI with more realistic thermal relaxation in the inner parts of PPDs, around \SI{5}{\AU} distance to the central star. 
By employing the thermal relaxation model by \citet{Malygin2017}, we were able to investigate how turbulence and structure formation are caused by the VSI, and how parameters like the radial stratification and the diffusion timescale act on the instability.
Our main results are:

\begin{itemize}
    \item The VSI can operate in the interior parts of PPDs, around \SI{5}{\AU}, \rev{under conditions obtained from our disk structure model. A disk with higher mass (and thus higher optical depth), might be less VSI-active, while a lower mass disk, with faster thermal relaxation respectively, might be more prone to VSI than our chosen disk structure.} 
    
    \item In the upper atmosphere of the cold regions of PPDs, in which the dust emission dominates over the gas emission, VSI is strongly hampered by the collisional dust-gas decoupling at low densities. We, thus, conclude that the thermal coupling of the dust and gas component is of great importance for the cooling time sensitive VSI in the regions beyond the water ice line. 
    
    \item The VSI reaches maximal turbulent stresses of $\alpha \sim 10^{-5}-10^{-3}$ at $\sim 2.5-3$ pressure scale heights above the midplane, depending on the radial stratification. 
    
    \item
    The mean vertical gas velocities reach a maximum at a height of $\approx 2.2$ pressure scale heights, with $v_z \approx \SI{0.04}{\kilo \meter \per \second}$.
    This result is consistent with the turbulent velocities obtained from turbulent line broadening \citep{Flaherty2015, Teague2016, Flaherty2017, Flaherty2018} and VSI turbulence could be taken into account for the interpretation of these observations.
    
    \item Our results show that the disk midplane is not laminar. The VSI zonal flows in fact cross the midplane with average velocities of $\approx \SIrange{0.01}{0.03}{\kilo \meter \per \second}$ depending on the disk's radial temperature gradient. This could influence the distribution of dust around the midplane.
     Dust might not be concentrated in the disk midplane, due to the stirring introduced by the VSI zonal flows.
    
    \item
     The growth rates of the instability linearly depend on the radial stratification, with $\Gamma P_0\sim 0.3$ for $\beta_T=-1$, as predicted by linear theory.
    
    \item Our three-dimensional simulation has shown the ability of the VSI to create long-lived anticyclonic vortices, that \rev{form} a central pressure maximum, even under non-ideal conditions in the inner parts of PPDs \rev{(a non-isothermal gas with finite cooling times)}.
    This process could be caused by the Rossby Wave Instability \citep[as seen in][]{Richard2016, Manger2018}. The vortices' longevity is possibly facilitated by the Subcritical Baroclinic Instability, which should be in operation due to the negative radial entropy gradient in our simulation \citep{Klahr2003, Petersen2007a, Petersen2007b}. 
    
    \item Anticyclonic vortices can survive over hundreds of orbits at $\sim \SI{5}{\AU}$ distance to the central star. They undergo little radial migration and mergers.
    The emerging vortices extend vertically over the whole VSI-active part of the simulation domain ($\pm 3$ pressure scale heights). In our simulations, vortices span $\approx 10$ pressure scale heights in azimuth and $\sim 1$ pressure scale height in radius. This is smaller than in the \SI{360}{\degree} simulations by \citet{Manger2018} and \citet{Manger2020}. Running azimuthally global simulations must, thus, be the next step, to check whether the smaller vortex sizes are caused by our smaller simulation domain, or by the different thermal relaxation regimes.
    
    \item It remains unclear how much the SBI contributes to the vortices' longevity. Future studies could assess its influence by probing different radial gradients in entropy.
    
    \item Modeling the dust evolution is a necessary next step to self-consistently simulate the dust-gas coupling and back reaction, the dust opacities, and thus the thermal relaxation times of the gas.
    An investigation of the dust evolution and accumulation in and around the vortices formed via VSI is necessary to assess the instability's influence on planetesimal formation.
    
    \item The influence of a more realistic vertical stratification has to be studied in future simulation, i.e. the impact of stellar irradiation, as seen in \citet{Flock2017, Flock2020}. As a next step it will be necessary to realize a simulation with a self-consistent vertical and radial stratification, where viscous heating and stellar irradiation lead to a complex temperature structure.
    
\end{itemize}

We conclude that the VSI is a robust mechanisms that leads to turbulence and structure formation in PPDs, even under non-ideal conditions \rev{like finite cooling times close to the midplane and shallow radial temperature gradients}.
The instability creates a complex distribution of turbulent stresses, depending on the local thermal relaxation timescales. The collisional coupling of the dust and gas component is of great importance for the emergence of the VSI in regions beyond the water ice line, and collisional decoupling sets a upper limit to the vertical extent of the VSI-active layer.

Especially the VSI's capability to form long-lived anticyclonic vortices, even in optically thick regions of PPDs, shows that this instability could be important in the formation process of planetesimals and planets.

\section*{Acknowledgments}
The authors thank the referee, Wladimir Lyra, for his
comments which helped improve the quality of this article.
T.P., H.K., and T.B. acknowledge the support of the German
Science Foundation (DFG) priority program SPP 1992
“Exploring the Diversity of Extrasolar Planets” under grant
Nos. BI 1816/7-2 and KL 1469/16-1/2. We would like to
thank the whole planet and star formation theory group of the
Max-Planck-Institute for Astronomy in Heidelberg for many
fruitful discussions of the topic and their help and advice. This
research was supported by the Munich Institute for Astro- and
Particle Physics (MIAPP) of the DFG cluster of excellence
“Origin and Structure of the Universe” and in part at KITP
Santa Barbara by the National Science Foundation under grant
No. NSF PHY17-48958. The authors gratefully acknowledge
the Gauss Centre for Supercomputing (GCS) for providing
computing time for a GCS Large-Scale Project (additional time
through the John von Neumann Institute for Computing (NIC))
on the GCS share of the supercomputer JUQUEEN \citep{Stephan2015} and now JEWELS at Jülich Supercomputing
Centre (JSC). GCS is the alliance of the three national
supercomputing centers HLRS (Universität Stuttgart), JSC
(Forschungszentrum Jülich), and LRZ (Bayerische Akademie
der Wissenschaften), funded by the German Federal Ministry
of Education and Research (BMBF) and the German State
Ministries for Research of Baden-Württemberg (MWK),
Bayern (StMWFK) and Nordrhein-Westfalen (MIWF). Additional simulations for the project of hydrodynamic instabilities
in PPDs were performed on the ISAAC cluster owned by the
MPIA and the COBRA and DRACO clusters of the Max Planck Society, both hosted at the Max Planck Computing and
Data Facility in Garching (Germany). Part of this work was
performed at the Aspen Center for Physics, which is supported
by National Science Foundation grant No. PHY-1607761.

\clearpage
\section{Appendix}
\subsection{The typical size of the VSI flow stucutures}

Our cooling time model relies on the assumption of a typical diffusion length scale in the optically thick parts of the PPD. 

To get an approximate value for the typical size of the VSI zonal flows, that we aim to reproduce in our simulation, we conduct a simulation of the inner parts of a disk with flux limited diffusion under the same conditions.
To that end, we utilize PLUTO's thermal conduction module, to solve the equation of flux limited ratiative diffusion \citep[FLD][]{Levermore1981}, instead of our Newtonian cooling model. 
This allows us to more realistically model the regions close to the midplane, and to determine what typical sizes VSI induced flow structures have, if thermal relaxation is caused by radiative diffusion.

\begin{figure*}[b]
    \centering
    \includegraphics[width=\textwidth]{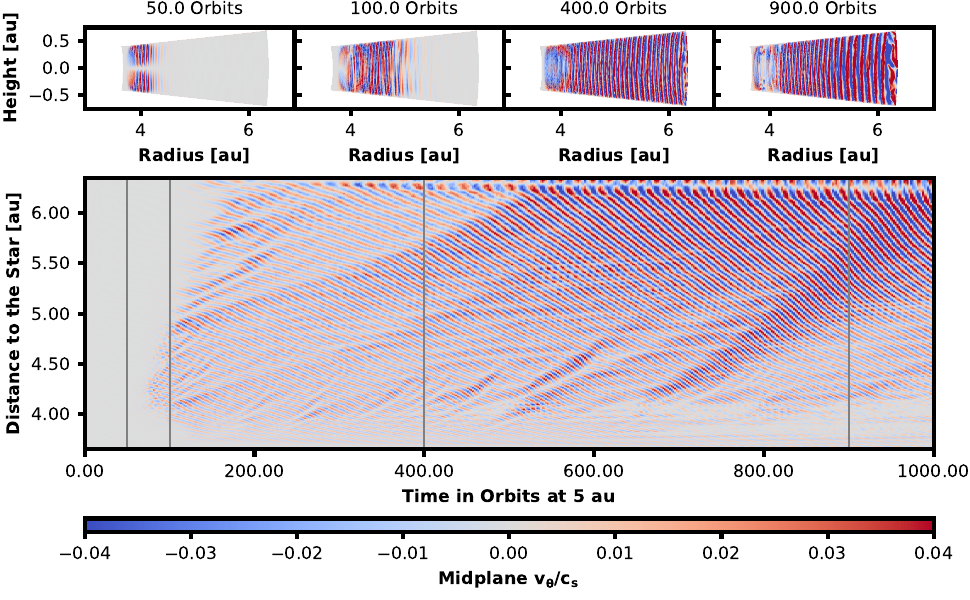}
    \caption{Time evolution of the vertical velocity in our simulation with FLD.}
    \label{fig:DiffVSI}
\end{figure*}

\begin{figure}[ht]
    \centering
    \includegraphics[width=0.45\textwidth]{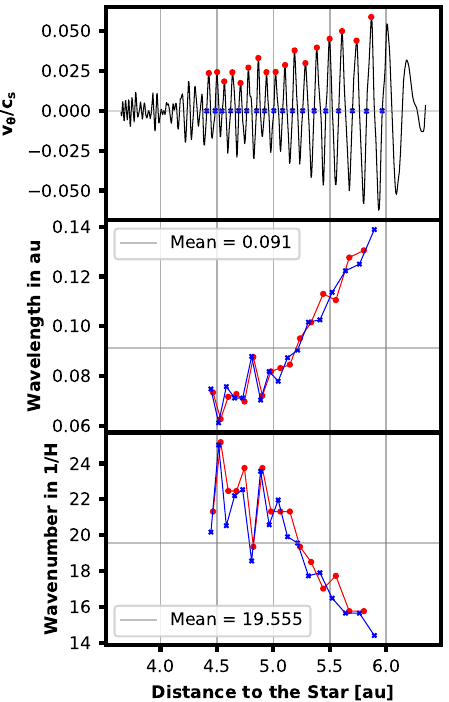}
    \caption{We determine the approximate size of the VSI zonal flows by measuring the mean distance of two consecutive changes of sign of the vertical velocity as depicted here (in blue). Alternatively, the distance between the maxima in velocity can be measured (in red), but we find the first method to produce less scatter.}
    \label{fig:wavenumber}
\end{figure}

\autoref{fig:DiffVSI} depicts the time evolution of the vertical velocities in this simulation. From visual inspection it is already evident that the size of the emerging flow structures does not significantly change over the run time of the simulation. 
In order to get the radial size of these zonal flows, we measure the distance between two consecutive changes in the sign of the vertical velocity, as shown in \autoref{fig:wavenumber}.

We find $kH\approx 20$, to be the average radial wavenumber in this simulation. 

\clearpage
\bibliography{Literature}

\end{document}